\documentclass[useAMS,usenatbib]{mn2e}
\usepackage{multirow}
\usepackage{epsfig}
\usepackage{subfigure}

\makeatletter
\let\jnl@style=\rm
\def\ref@jnl#1{{\jnl@style#1}}

\def\aj{\ref@jnl{AJ}}                   
\def\araa{\ref@jnl{ARA\&A}}             
\def\apj{\ref@jnl{ApJ}}                 
\def\apjl{\ref@jnl{ApJ}}                
\def\apjs{\ref@jnl{ApJS}}               
\def\ao{\ref@jnl{Appl.~Opt.}}           
\def\apss{\ref@jnl{Ap\&SS}}             
\def\aap{\ref@jnl{A\&A}}                
\def\aapr{\ref@jnl{A\&A~Rev.}}          
\def\aaps{\ref@jnl{A\&AS}}              
\def\azh{\ref@jnl{AZh}}                 
\def\baas{\ref@jnl{BAAS}}               
\def\cjaa{\ref@jnl{ChJAA}}		
\def\jrasc{\ref@jnl{JRASC}}             
\def\memras{\ref@jnl{MmRAS}}            
\def\mnras{\ref@jnl{MNRAS}}             
\def\nar{\ref@jnl{NewAR}}               
\def\na{\ref@jnl{NewA}}                 
\def\pra{\ref@jnl{Phys.~Rev.~A}}        
\def\prb{\ref@jnl{Phys.~Rev.~B}}        
\def\prc{\ref@jnl{Phys.~Rev.~C}}        
\def\prd{\ref@jnl{Phys.~Rev.~D}}        
\def\pre{\ref@jnl{Phys.~Rev.~E}}        
\def\prl{\ref@jnl{Phys.~Rev.~Lett.}}    
\def\pasp{\ref@jnl{PASP}}               
\def\pasj{\ref@jnl{PASJ}}               
\def\qjras{\ref@jnl{QJRAS}}             
\def\skytel{\ref@jnl{S\&T}}             
\def\solphys{\ref@jnl{Sol.~Phys.}}      
\def\sovast{\ref@jnl{Soviet~Ast.}}      
\def\ssr{\ref@jnl{Space~Sci.~Rev.}}     
\def\zap{\ref@jnl{ZAp}}                 
\def\nat{\ref@jnl{Nature}}              
\def\iaucirc{\ref@jnl{IAU~Circ.}}       
\def\aplett{\ref@jnl{Astrophys.~Lett.}} 
\def\apspr{\ref@jnl{Astrophys.~Space~Phys.~Res.}}
\def\bain{\ref@jnl{Bull.~Astron.~Inst.~Netherlands}}
\def\fcp{\ref@jnl{Fund.~Cosmic~Phys.}}  
\def\gca{\ref@jnl{Geochim.~Cosmochim.~Acta}}   
\def\grl{\ref@jnl{Geophys.~Res.~Lett.}} 
\def\jcp{\ref@jnl{J.~Chem.~Phys.}}      
\def\jgr{\ref@jnl{J.~Geophys.~Res.}}    
\def\jqsrt{\ref@jnl{J.~Quant.~Spec.~Radiat.~Transf.}}
\def\memsai{\ref@jnl{Mem.~Soc.~Astron.~Italiana}}
\def\nphysa{\ref@jnl{Nucl.~Phys.~A}}   
\def\physrep{\ref@jnl{Phys.~Rep.}}   
\def\physscr{\ref@jnl{Phys.~Scr}}   
\def\planss{\ref@jnl{Planet.~Space~Sci.}}   
\def\procspie{\ref@jnl{Proc.~SPIE}}   

\makeatother

\title[{\it {\it INTEGRAL}} Narrow Line Seyfert 1]{Narrow Line Seyfert 1 galaxies at hard X-rays\thanks{Based on observations obtained with {\it INTEGRAL}/IBIS,
XMM-\textit{Newton} and {\it Swift}/XRT}}
\author[Francesca Panessa et al.]{F. Panessa$^1$\thanks{E-mail: francesca.panessa@iasf-roma.inaf.it}, A. De Rosa$^1$, L. Bassani$^2$, 
A. Bazzano$^1$, A. Bird$^3$, R. Landi$^2$, 
\newauthor A. Malizia$^2$, G. Miniutti$^4$, M. Molina$^5$, P. Ubertini$^1$ \\
$^1$Istituto di Astrofisica Spaziale e Fisica Cosmica (IASF-INAF), via del Fosso del Cavaliere 100, 00133 Roma, Italy\\
$^2$Istituto di Astrofisica Spaziale e Fisica Cosmica (IASF-INAF), Via P. Gobetti 101, 40129 Bologna, Italy\\
$^3$School of Physics and Astronomy, University of Southampton, Southampton, SO17 1BJ, UK\\ 
$^4$Centro de Astrobiologia (CSIC-INTA), PO Box 78, 28691, Villanueva de la Ca\~nada, Madrid, Spain \\
$^5$Istituto di Astrofisica Spaziale e Fisica Cosmica (IASF-INAF), Via Bassini 15, 20122 Milano,Italy\\
}
\begin{document}

\date{}

\pagerange{\pageref{firstpage}--\pageref{lastpage}} \pubyear{2002}

\maketitle

\label{firstpage}

\begin{abstract}

Narrow Line Seyfert 1 (NLSy1) galaxies are a peculiar class of type 1 AGN (BLSy1). The X-ray properties
of individual objects belonging to this class are often extreme and associated with accretion at high Eddington ratios.
Here we present a study on a sample of 14 NLSy1 galaxies selected at hard X-rays ($>$ 20 keV) from the fourth {\it INTEGRAL}/IBIS catalogue.
The 20-100 keV IBIS spectra show hard X-ray photon indeces flatly
distributed ($\Gamma_{20-100 keV}$ ranging from $\sim$ 1.3 to $\sim$ 3.6) with an average value of $<$$\Gamma_{20-100 keV}$$>$ $=$ 2.3$\pm$0.7,
compatible with a sample of hard X-ray BLSy1 average slope. Instead, NLSy1 show steeper spectral indeces with respect to BLSy1
when broad-band spectra are considered. Indeed, we combine XMM-\textit{Newton} and {\it Swift}/XRT with {\it INTEGRAL}/IBIS data sets 
to obtain a wide energy spectral coverage (0.3-100 keV). A constraint on the high energy cut-off and on the reflection component
is achieved only in one source, Swift~J2127.4+5654 (E$_{cut-off}$ $\sim$ 50 keV, R=1.0$^{+0.5}_{-0.4}$). 
Hard X-ray selected NLSy1 do not display particularly strong soft excess emission, while absorption fully or partially covering the continuum is often measured
as well as Fe line emission features. Variability is a common trait in this sample, both at X-ray and at hard X-rays.
The fraction of NLSy1 in the hard X-ray sky is likely to be $\sim$ 15\%, in agreement with estimates derived in optically selected NLSy1 samples. 
We confirm the association of NLSy1 with small black hole masses with a peak at 10$^{7}$ M$_{\odot}$ in the distribution, 
however hard X-ray NLSy1 seem to occupy the lower tail of the Eddington ratios distribution of classical NLSy1. 

\end{abstract}


\section{Introduction}

\begin{table*}
\label{sam}
\caption{\bf The hard X-ray selected sample of Narrow Line Seyfert 1 galaxies}
\begin{center}
\begin{tabular}{lrrrccccccccc}
\hline
\hline
\multicolumn{1}{c}{Name} &
\multicolumn{1}{c}{RA} &
\multicolumn{1}{c}{DEC} &
\multicolumn{1}{c}{z} &
\multicolumn{1}{c}{H$_{\beta}$} &
\multicolumn{1}{c}{[O\textsc{iii}]/H$_{\beta}$}  &
\multicolumn{1}{c}{[FeII]/H$_\beta$}  &
\multicolumn{1}{c}{N$_{H, Gal}$}  &
\multicolumn{1}{c}{M$_{BH}$}  &
\multicolumn{1}{c}{Ref}  &
\multicolumn{1}{c}{S$_{R}$}  &
\multicolumn{1}{c}{Ref}  \\
\multicolumn{1}{c}{(1)} &
\multicolumn{1}{c}{(2)} &
\multicolumn{1}{c}{(3)} &
\multicolumn{1}{c}{(4)} &
\multicolumn{1}{c}{(5)} &
\multicolumn{1}{c}{(6)} &
\multicolumn{1}{c}{(7)} &
\multicolumn{1}{c}{(8)} &
\multicolumn{1}{c}{(9)} &
\multicolumn{1}{c}{(10)} &
\multicolumn{1}{c}{(11)} &
\multicolumn{1}{c}{(12)}  \\
\hline
\hline
1H~0323+342	    & 03 24 41.2 & +34 10 46  & 0.0610 & 1650 & $\sim$0.12    & 2.0	      & 14.5 & 7.0  & 1	    &  362 & 1    \\
NGC~4051	    & 12 03 09.6 & +44 31 53  & 0.0023 & 1120 & 0.55	      & 0.25	      & 1.30 & 0.17 & 2,3   &  9.4 & 2	    \\
Mrk~766		    & 12 18 26.5 & +29 48 46  & 0.0129 & 1630 & 1.83	      & 0.35	      & 1.70 & 0.43 & 2,4   &  20.5 & 2	    \\  
NGC~4748	    & 12 52 12.4 & -13 24 53  & 0.0146 & 1565 & 1.34	      & 0.55	      & 3.60 & 0.42 & 2,4   &  6.0 & 2		   \\  
Mrk~783		    & 13 02 58.8 & +16 24 27  & 0.0672 & 1655 & 2.29	      & $<$0.5        & 2.00 & 1.43 & 2,4   & 32.9 & 3 	   \\
NGC~5506	    & 14 13 14.9 & -03 12 27  & 0.0061 &  -   & 7.5	      & -	      & 3.81 & 0.2  & 5,6   & 227.2 & 2	   \\
IGR~J14552-5133     & 14 55 17.8 & -51 34 17  & 0.0160 & 1700 & 0.70$\pm$0.06 & $\leq$ 1.70   & 33.7 & 0.2  & 7     & $-$ &$-$ 	     \\
IRAS~15091-2107     & 15 11 59.8 & -21 19 02  & 0.0446 & 1480 & 1.0	      & 0.70	      & 8.40 & 1.0  & 8     & 12.0 & 4	     \\
IGR~J16185-5928     & 16 18 25.7 & -59 26 46  & 0.0350 & 4000 & 0.20$\pm$0.03 & 0.70$\pm$0.06 & 24.7 & 2.8  & 7     & $-$ &$-$	    \\  
IGR~J16385-2057     & 16 38 30.9 & -20 55 25  & 0.0269 & 1700 & 0.50$\pm$0.06 & 1.20$\pm$0.20 & 12.0 & 0.7  & 9    & 6.8 & 3	    \\   
IGR~J16426+6536     & 16 43 04.1 & +65 32 51  & 0.3230 & 2100 & 0.41$\pm$0.04 & 0.85	      & 3.00 & 1.1  & 10    & $-$ &$-$	    \\
IGR~J19378-0617     & 19 37 39.0 & -06 13 06  & 0.0106 & 2700 & 0.52$\pm$0.04 & 0.89$\pm$0.06 & 15.0 & 0.3  & 11    &  42.2     & 3  \\
ESO~399-IG 020      & 20 06 58.1 & -34 32 58  & 0.0249 & 2425 & -	      & -	      & 7.10 & -    & 12    & 8.4  & 3 		   \\
Swift~J2127.4+5654  & 21 27 45.6 & +56 56 36  & 0.0140 & 2000 & 0.72$\pm$0.05 & 1.30$\pm$0.20 & 78.7 & 1.5  & 13,14 & 6.4 & 3 	  \\
\hline
\end{tabular}
\end{center}
Notes: (1): Galaxy name; col (2)-(3) Optical position in epoch J2000; (4) Redshift; (5) FWHM of H$_{\beta}$ (km/s); (6): [O\textsc{iii}]/H$_{\beta}$ ratio; 
(7): [FeII]/H$_\beta$ ratio; (8): Galactic column density in units of 10$^{20}$ cm$^{-2}$; (9): Black hole mass in units
of 10$^{7}$ solar masses; (10): References for optical and black hole mass measurements: 
(1) Zhou et al. (2007); (2) V{\'e}ron-Cetty, M.-P., \& V{\'e}ron, P. (2006);
(3) Denney et al. (2009); (4) Wang \& Lu (2001); (5) Nagar et al. (2002); (6) Nikolaiuk et al. (2009);
(7) Masetti et al. (2006); (8) Osterbrock \& De Robertis (1985); 
(9) Masetti et al. (2008); (10) Masetti et al. (2009); (11) Rodriguez-Ardila et al. (2000); (12) Dietrich et al. (2005);
(13) Malizia et al. (2008); (14) Halpern 2006. (11): Radio fluxes in mJy; (12): References for radio fluxes:
(1) 5~GHz, Becker et al. (1991); (2) 5~GHz, Gallimore et al. (2006); (3) 1.4~GHz, Condon et al. (1998); (4) 5~GHz, Ulvestad et al. (1995).
\end{table*}

Narrow Line Seyfert 1 (NLSy1) are AGN with
optical spectral properties similar to those of Seyfert 1 galaxies (Broad Line Seyfert 1, BLSy1), 
except for having narrow Balmer lines and strong optical FeII lines (Osterbrock \& Pogge 1985).
The classical definition of NLSy1 (Goodrich 1989) relies on three simple
criteria: (i) the full width half maximum (FWHM) of the H$_\beta$ line $<$ 2000 km s$^{-1}$; 
(ii) the [OIII] $\lambda$5007/H$_\beta$ ratio $<$3; and (iii) unusually
strong FeII. The NLSy1 and BLSy1 classes are not completely distinct, 
for instance V{\'e}ron-Cetty \& V{\'e}ron (2006) noticed 
a continuity between their optical spectral properties.

The most extreme characteristics of NLSy1 are seen in the X-ray domain (Gallo 2006). A strong and variable soft excess
emission below $\sim$1 keV (George et al. 2000, Boller et al. 1996, Turner et al. 1999) is more frequently present 
than in BLSy1 (Leighly 1999). The 2-10 keV spectral slope is usually steeper in NLSy1 with respect to BLSy1 (Leighly 1999,
Brandt et al. 1997) and a sharp spectral drop at about 7 keV has been observed in some objects (Uttley et al. 2004,
Longinotti et al. 2003, Boller et al. 2003, Boller et al. 2002). The complex NLSy1 X-ray spectrum has been explained in terms of
relativistic blurred disc reflection (e.g., Fabian et al. 2004) or ionized or neutral
absorption either totally or partially covering the X-ray source (e.g., Gierli{\'n}ski  \& Done 2004, Tanaka et al. 2004).

A possible interpretation for the peculiar observational properties of NLSy1 is that these systems
are accreting close to their Eddington limit, suggesting that, compared 
to their BLSy1 counterparts, they should host smaller black hole masses (M$_{BH}$ $<$ 10$^{8}$ M$_{\odot}$), 
as measured in the majority of the sources (Wandel et al. 1999, Grupe \& Mathur 2004, Whalen et al. 2006). 
Under the assumption that the H$_{\beta}$ emitting region is gravitationally bounded to the central black hole,
small black hole masses help in explaining the narrow emission lines observed in these objects.
It has been shown that NLSy1 galaxies seem not to follow the M$_{BH}$-$\sigma$ relation (Komossa \& Xu 2007, Grupe \& Mathur 2004),
suggesting that they may be AGNs in their early stage of development, where small black hole masses
grow through accretion in already formed bulges with strong implications for the black hole versus galaxy evolution (Grupe \& Mathur 2004, Mathur et al. 2001). 
However, lately, Marconi et al. (2008) suggested that radiation pressure due to ionizing photons could significantly 
affect virial black hole mass estimates in highly accreting objects such as NLSy1, resulting in an underestimate of their masses.

The interest on NLSy1 has further increased due to the detection of gamma-ray emission
from four NLSy1 with {\it Fermi} (Abdo et al. 2009) suggesting that NLSy1 may also form
a powerful radio jet and at the same time be powered by black holes of moderate masses
(Yuan et al. 2008).

The hard X-rays is an energy range where NLSy1 have been poorly investigated so far,
only studies on sparse objects are available. {\it BeppoSAX} observations of a small sample of
NLSy1 detected hard X-ray emission in only 2 out of 7 sources (Comastri 2000). 
{\it Suzaku} is now providing detailed broad-band studies of targeted NLSy1 (e.g., Ponti et al. 2010, Miniutti et al. 2009, Terashima et al. 2009). Recent hard X-ray surveys performed by the IBIS instrument 
(Ubertini et al. 2003) on board {\it INTEGRAL} (Winkler et al. 2003) and the Burst Alert Telescope (BAT, Barthelmy et al. 2005) on board {\it Swift} (Geherels et al. 2004) are imaging
the hard X-ray sky providing a number of sources detected for the first time at these energies
with half of them being newly discovered (Malizia et al. 2010, Landi et al. 2010, Tueller et al. 2010,
Tueller et al. 2009, Ajello et al. 2008, Winter et al. 2008, Bassani et al. 2006). 

Indeed, the fourth IBIS survey (Bird et al. 2010) has allowed the detection of more than 700 
hard X--ray sources over the whole sky above 20 keV, down to an average flux level 
of about 1 mCrab and with positional accuracies ranging between 0.2 and 
$\sim$5 arcmin, depending on the source strength. Among the detected AGN by IBIS, a small sample of NLSy1 is identified.
A first analysis of the hard X-ray properties of a sample
of five NLSy1 listed in the third IBIS catalogue (Bird et al. 2007) has been presented in Malizia et al. (2008),
combining {\it Swift}/XRT (X-ray Telescope, Burrows et al. 2005) and {\it INTEGRAL}/IBIS spectra.

In this work, we present the broad-band X-ray properties of a sample of 14 NLSy1
detected by {\it INTEGRAL} and present in the fourth IBIS catalogue (Bird et al. 2010).
New XMM-\textit{Newton} datasets for 5 NLSy1 of the sample
have been combined with {\it INTEGRAL} data to obtain a wide energy spectrum up to $\sim$ 100 keV. In 5 other objects,
XRT data are available below 10 keV, and, for IGR J16426+6536, only an XRT detection is available.
Published best-fit spectral parameters for NGC~4051, Mrk~766 and NGC~5506 were taken from the most recent X-ray 
dedicated works (Terashima et al. 2009, Turner et al. 2007 and Guainazzi et al. 2010, respectively).

The paper is organized in the following manner: Section 2 describes the sample; in Section 3 the
observation and data reduction procedure of the {\it INTEGRAL}, XMM-\textit{Newton} and XRT datasets is reported;
Section 4 illustrates the broad-band XMM-\textit{Newton} and {\it INTEGRAL} spectral fitting results;
Section 5 is devoted to the discussion of the fraction of NLSy1 in hard X-ray samples and of the
black hole mass, Eddington ratios and radio loudness distributions. Finally, in Section 6 we report our conclusions.
Throughout this paper we assume a flat $\Lambda$CDM cosmology with ( $\Omega_{\rm M}$,  $\Omega_{\rm\Lambda}) = (0.3$,
0.7) and a Hubble constant of 70 km s$^{-1}$ Mpc$^{-1}$ (Bennett et al. 2003).

\begin{table*}
\begin{center}
\caption{\bf The hard X-ray data}
\label{table=hard}
\begin{tabular}{lrrcccccc}
\hline
\hline
\multicolumn{1}{c}{Name} &
\multicolumn{1}{c}{Significance}  &
\multicolumn{1}{c}{Exposure} &
\multicolumn{1}{c}{Map Code} &
\multicolumn{1}{c}{$\Gamma_{\rm 20-100 keV}$} &
\multicolumn{1}{c}{$\chi^{2}$/dof} &
\multicolumn{1}{c}{F$_{\rm 20-100 keV}$} &
\multicolumn{1}{c}{L$_{\rm 20-100 keV}$} &
\multicolumn{1}{c}{F$_{\rm {\it Swift}/BAT}$} \\
\multicolumn{1}{c}{(1)} &
\multicolumn{1}{c}{(2)} &
\multicolumn{1}{c}{(3)} &
\multicolumn{1}{c}{(4)} &
\multicolumn{1}{c}{(5)} &
\multicolumn{1}{c}{(6)} &
\multicolumn{1}{c}{(7)} &
\multicolumn{1}{c}{(8)} &
\multicolumn{1}{c}{(9)}  \\
\hline
\hline
1H 0323+342	    		&  6.8 & 128 & 18-60  & 1.55$^{+0.80}_{-1.20}$   & 1.28/5 &  5.3 & 44.67 & 1.7  \\	    
NGC 4051	        	& 10.0 & 528 & 20-40  & 2.16$^{+0.41}_{-0.38}$   & 7.59/8 &  3.6 & 41.62 & 2.3  \\
Mrk 766		        	&  7.1 & 584 & 20-40  & 2.85$^{+0.85}_{-0.68}$	 & 6.00/7 &  2.0 & 42.88 & 1.3  \\  
NGC 4748	        	&  5.2 & 841 & 18-60  & 1.43$^{+0.92}_{-1.05}$   & 9.05/7 &  1.7 & 42.91 & 0.6  \\  
Mrk 783		        	&  5.8 & 626 & 20-100 & 3.62$^{+2.00}_{-1.54}$   & 3.21/7 &  0.8 & 43.97 & 1.1  \\	     
NGC 5506	    		&  9.0 &25.1 & 18-60  & 2.41$^{+0.68}_{-0.59}$   & 7.64/6 & 16.1 & 43.12 & 14.8 \\
IGR J14552-5133         	&  8.8 &2558 & 18-60  & 1.90$^{+0.44}_{-0.46}$   & 5.40/8 &  1.5 & 42.93 & 1.3  \\
IRAS 15091-2107         	&  4.5 & 258 & 20-40  & 2.95$^{+1.13}_{-0.92}$   & 2.59/9 &  1.8 & 43.94 & $-$  \\
IGR J16185-5928         	&  8.5 &2213 & 18-60  & 1.34$^{+0.82}_{-0.40}$   & 3.26/8 &  1.7 & 43.69 & 1.0  \\
IGR J16385-2057         	&  5.5 &1066 & 20-100 & 3.07$^{+1.00}_{-0.71}$   & 5.47/6 &  1.4 & 43.38 & 1.3  \\
IGR J16426+6536         	&  6.2 &  97 & 18-60  & 3.34$^{+1.98}_{-1.57}$   & 2.07/5 &  2.5 & 46.10 & $-$  \\
IGR J19378-0617         	&  6.1 & 727 & 20-100 & 2.14$^{+0.65}_{-0.56}$   &18.66/10&  1.8 & 42.66 & 1.4  \\
ESO 399-IG 020          	&  5.5 & 932 & 20-100 & 1.54$^{+0.63}_{-0.58}$   & 1.90/8 &  1.7 & 43.39 & 1.3  \\
Swift J2127.4+5654      	& 19.4 &1781 & 18-60  & 2.50$^{+0.24}_{-0.24}$   & 7.21/7 &  3.4 & 43.17 & 2.2  \\
\hline		    			  
\end{tabular}
\end{center}
Notes: (1): Galaxy name; (2): Maximum source significance in the map; (3): On-source exposure (ks);
(4): Energy range with source maximum significance; (5): hard X-ray photon index; (6): Chi squared and degree of freedom ratio;
(7): 20-100 keV {\it INTEGRAL}/IBIS flux in 10$^{-11}$ ergs cm$^{-2}$ s$^{-1}$; 
(8): Logarithm of 20-100 keV {\it INTEGRAL}/IBIS luminosity in ergs s$^{-1}$; (9): 20-100 keV {\it Swift}/BAT flux in 10$^{-11}$ ergs cm$^{-2}$ s$^{-1}$ adapted from Cusumano et al. (2010).
\end{table*}

\section{The Narrow Line Seyfert 1 sample}

In Table~\ref{sam} we present the sample of 14 hard X-ray selected NLSy1
from the fourth IBIS/ISGRI soft gamma-ray survey catalog
(see Bird et al. 2010 for a complete description of hard X-ray source detection and identification). 
In the following, we discuss our check on the optical classification for each source.
1H 0323+342 has been classified as a NLSy1 by Zhou et al. (2007), 
reporting strong optical Fe II complexes. From a multiwavelength study, this object shows 
the dual properties of both a NLSy1 and a blazar. Indeed, it has been recently
detected by the {\it Fermi}/LAT instrument (Abdo et al. 2009). 
The optical spectral data of NGC~4051, Mrk~766, NGC~4748 and Mrk~783 have been
taken from V{\'e}ron-Cetty \& V{\'e}ron (2006) (see this work for detailed references on each source). 
NGC~5506 is reported as a type 1.9 Seyfert galaxy in Bird et al. (2010), however its optical classification
has been largely debated. Nagar et al. (2002)
proposed a NLSy1 classification based on the permitted OI line (with FHWM $<$ 2000 km s$^{-1}$) 
and on the 1 $\mu$m Fe II line observed between 0.9-1.4 $\mu$m. 
The optical classification of IGR~J14552-5133, IGR~J16185-5928, IGR~J16385-2057,
IGR~J19378-0617 and SWIFT~J2127.4+5654 has been extensively discussed in Malizia et al. (2008)
where the relative references are reported\footnote{These sources have been classified during 
the optical follow-up campaigns which are on-going to identify the {\it INTEGRAL}/IBIS sources 
(Masetti et al. 2010 and references therein). See the web page 
{\tt http://www.iasfbo.inaf.it/extras/IGR/main.html} for the information about the {\it 
INTEGRAL} sources identified using optical or near-infrared (NIR) observations.}.

The optical classification of NLSy1 is not always straightforward.
Sulentic et al. (2000) and, lately Veron et al. (2006), pointed out that
many AGN with H$_\beta$ $>$ 2000 km s$^{-1}$ behave like NLSy1 (e.g., strong FeII, strong and variable soft X-ray excess,
etc.) and have identified the ratio FeII/H$_\beta$ $>$ 0.5 as a more appropriate 
dividing line between BLSy1 and NLSy1. 
This is the case of IGR J16185-5928 and IGR J19378-0617 with H$_\beta$ $>$ 2000 km s$^{-1}$ but 
classified as NLSy1 based on their 
[OIII] $\lambda$5007/H$_\beta$ and FeII/H$_\beta$ ratios.  
Also in ESO 399-IG 020, the H$_\beta$ FWHM is slightly above 2000 km/s and the source was 
originally reported as a type 1 Seyfert in Zhou \& Wang (2002); lately the source has been reclassified 
as NLSy1 by Dietrich et al. (2005) (although the authors do not report emission line ratios but only the 
H$_{\beta}$ FWHM and its flux, i.e., 1.2 $\times$ 10$^{-13}$ ergs cm$^{-2}$ s$^{-1}$).
In Osterbrock \& De Robertis (1985), IRAS 15091-2107 
is classified as a type 1 Seyfert, notwithstanding that the HI emission lines are only slightly broader that the forbidden lines
and FeII emission is as strong as observed in NLSy1. Subsequently,
Goodrich (1989) classified this source as a NLSy1.
The classification as a NLSy1 for IGR~J16426+6536 is given in Masetti et al. (2009), while  
Butler et al. (2009) classified IGR~J16426+6536 as a Seyfert 1.5. However, the two observations were separated
by several months suggesting possible intrinsic variability.

References to black hole mass estimates are reported in Table~\ref{sam}, all of them were calculated
through the radius-luminosity relation as in Kaspi et al. (2000). In Table~\ref{sam}, we also report radio fluxes
in mJy and their relative references.

\section{Observations and data reduction}

\begin{table*}
\caption{\bf XMM-\textit{Newton} and {\it Swift}/XRT Data Observation Details}
\small{
\begin{center}
\begin{tabular}{lcccc}
\hline
\hline
\multicolumn{1}{c}{Name} &
\multicolumn{1}{c}{Instrument} &
\multicolumn{1}{c}{Obs. Date} &
\multicolumn{1}{c}{Exposure} &
\multicolumn{1}{c}{Filter} \\
\multicolumn{1}{c}{(1)} &
\multicolumn{1}{c}{(2)} &
\multicolumn{1}{c}{(3)} &
\multicolumn{1}{c}{(4)} &
\multicolumn{1}{c}{(5)} \\
\hline
\hline
1H 0323+342	    & XRT & 2006-07-05 & 8071 & $-$  \\
		    & XRT & 2006-07-09 & 8802 & $-$  \\
                    & XRT & 2007-11-04 & 1885 & $-$  \\
IRAS 15091-2107     & XMM & 2005-07-26 & 9305 & medium \\
NGC4748		    & XRT & 2005-12-28 & 2243 & $-$  \\
		    & XRT & 2007-01-09 & 2266 & $-$  \\
Mrk 783		    & XRT & 2008-05-09 & 6305 & $-$  \\
		    & XRT & 2008-08-16 & 4487 & $-$  \\
IGR J14552-5133     & XRT & 2006-12-27 & 3728 & $-$  \\
     		    & XRT & 2007-01-02 & 4695 & $-$  \\
IGR J16185-5928     & XMM & 2009-02-18 & 8402 & thin     \\  
IGR J16385-2057     & XMM & 2009-03-15 & 7807 & medium  \\
		    & XMM & 2010-02-15 & 19800& medium \\   
IGR J19378-0617     & XMM & 2009-04-28 & 12240 & thin  \\
ESO 399-IG 020	    & XRT & 2010-08-27 & 8361 & $-$ \\
Swift J2127.4+5654  & XMM & 2009-11-11 & 8631 & thin\\
\hline
\end{tabular}
\end{center}
Notes: (1): Galaxy name; (2): Instrument; (3): Observation date;
(4): pn observation exposures in sec; (5): pn filters.} 
\label{table=obs_info}
\end{table*}

\subsection{The INTEGRAL data}\label{ibis}

The {\it INTEGRAL}/IBIS data used in this work have been collected in the 4th IBIS survey (Bird et al. 2010) consisting of 7355
pointings (called Science Windows, ScW, lasting about 2000 sec each) performed from the beginning of the Mission  
in November 2002 up to April 2008 and including both all available public and Core Programme data.  
IBIS/ISGRI images for each available pointing were generated in various energy bands using the ISDC
off-line scientific analysis software OSA (Goldwurm et al. 2003). Then all images have been mosaicked to create
significance map at revolution level (each orbit lasting about 3 days) and for all the available pointings. Count rates at the
position of the source were extracted from individual images in order to provide light curves in various energy bands; from these
light curves, average fluxes were then extracted and combined to produce an average source spectrum (see Bird et al. 2010 for
details). 

A spectral fit with a simple power-law model has been performed using the IBIS spectra.
In Table~\ref{table=hard} we report the source significance, the total on-source exposure, 
the map code as from Bird et al. (2010), indicating the best significance energy band, 
the hard X-ray photon index, the fit chi-squared over degrees of freedom ratio,
the 20-100 keV flux and luminosity. We also report the 20-100 keV flux from the 
{\it Swift}/BAT 54 months catalogue, estimated from the averaged 14-150 keV fluxes listed in Cusumano et al. (2010), 
assuming $\Gamma$ = 2.

\begin{figure*}
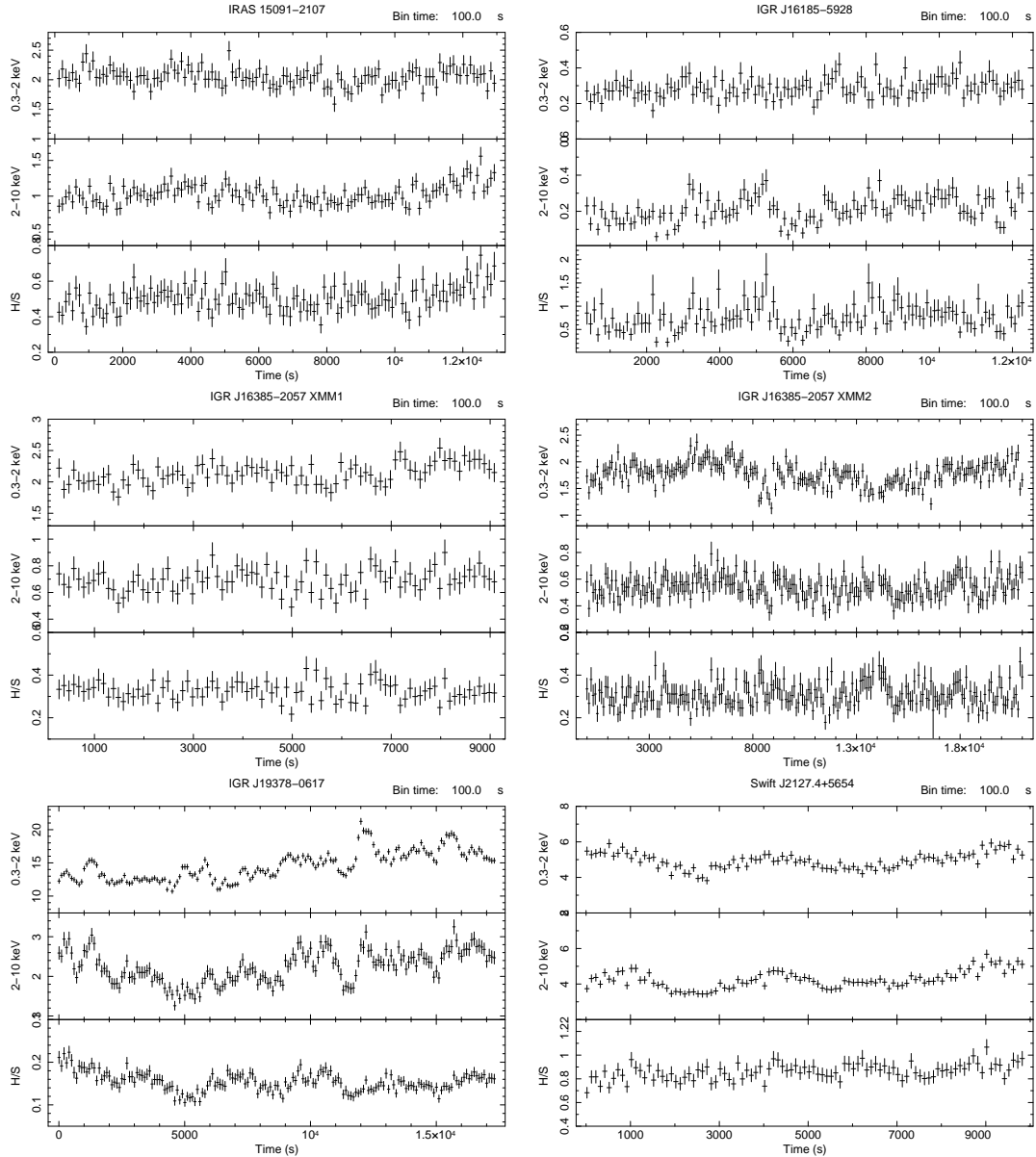

\begin{center}
\parbox{16cm}{
\includegraphics[width=0.3\textwidth,height=0.3\textheight,angle=-90]{IRAS_lc.ps}
\includegraphics[width=0.3\textwidth,height=0.3\textheight,angle=-90]{16185_lc.ps}}
\parbox{16cm}{
\includegraphics[width=0.3\textwidth,height=0.3\textheight,angle=-90]{16385_lc_xmm1.ps}
\includegraphics[width=0.3\textwidth,height=0.3\textheight,angle=-90]{16385_lc_xmm2.ps}}
\parbox{16cm}{
\includegraphics[width=0.3\textwidth,height=0.3\textheight,angle=-90]{19378_lc.ps}
\includegraphics[width=0.3\textwidth,height=0.3\textheight,angle=-90]{lc_swift.ps}}
\caption{XMM-\textit{Newton} light curves}
\label{figure=lcurve}
\end{center}
\end{figure*}

\begin{table*}
\footnotesize{
\caption{\bf XMM-\textit{Newton} light curve analysis.}
\label{table=lc}
\begin{center}
\begin{tabular}{lcrcrcr}
\hline
\hline
\multicolumn{1}{c}{Name} &
\multicolumn{1}{c}{Count rate (0.3-2 keV)} &
\multicolumn{1}{c}{$\chi^{2}$/dof} &
\multicolumn{1}{c}{Count rate (2-10 keV)} &
\multicolumn{1}{c}{$\chi^{2}$/dof} &
\multicolumn{1}{c}{Hardness ratio} &
\multicolumn{1}{c}{$\chi^{2}$/dof}\\
\multicolumn{1}{c}{(1)} &
\multicolumn{1}{c}{(2)} &
\multicolumn{1}{c}{(3)} &
\multicolumn{1}{c}{(4)} &
\multicolumn{1}{c}{(5)} &
\multicolumn{1}{c}{(6)} &
\multicolumn{1}{c}{(7)}   \\
\hline
\hline
IRAS 15091-2107                & 2.03$\pm$0.02 & {\bf 130/129} 	& 1.01$\pm$0.01 & 226/129 	& 0.49$\pm$0.02 & 162/129 \\
IGR J16185-5928                & 0.28$\pm$0.01 & {\bf 106/120} 	& 0.18$\pm$0.01 & 324/120 	& 0.61$\pm$0.02 & 190/120 \\
IGR J16385-2057 XMM1           & 2.12$\pm$0.02 & 107/89 	& 0.68$\pm$0.01 & {\bf 91/89}   & 0.32$\pm$0.01 & {\bf 77/89}\\
IGR J16385-2057 XMM2           & 1.67$\pm$0.02 & 1954/209 	& 0.49$\pm$0.01 & 817/209 	& 0.29$\pm$0.01 & 254/209\\
IGR J19378-0617                &14.25$\pm$0.05 & 5421/174 	& 2.16$\pm$0.02 & 1335/174 	& 0.15$\pm$0.01 &  662/174\\
Swift J2127.4+5654     	       & 4.89$\pm$0.03 & 401/99	 	& 4.18$\pm$0.03 & 532/99 	& 0.85$\pm$0.02 & 138/99 \\
\hline				
\end{tabular}
\end{center}
Note: (1): Galaxy name; (2): Count rate (counts s$^{-1}$) in the 0.3-2 keV band; (3): Chi-squared and degrees of freedom from fitting
the 0.3-2 keV light curve with a constant; (4): Count rate (counts s$^{-1}$) in the 2-10 keV band; (5): 
Chi-squared and degrees of freedom from fitting the 2-10 keV light curve with a constant. (6): Hardness ratio (2-10 keV/0.3-2 keV);
(7): Chi-squared and degrees of freedom from fitting the hardness ratio.
In bold face the $\chi^{2}$/dof with variability at less than 3 sigma confidence level are marked.}
\end{table*}

\subsection{The XMM-\textit{Newton} data}

In Table~\ref{table=obs_info} the log of XMM-\textit{Newton} observations 
is listed. The software SASv9.0 has been used
for the data reduction procedure. In many cases, observations were heavily affected from high-background flares,
therefore data were cleaned through an inspection of the high energy light curves, effective exposures obtained after screening
are those listed in Table~\ref{table=obs_info}. The source spectra and light curves
were extracted from circular regions centered on the sources while background
circular regions were chosen to be free from contaminating sources. 
The {\it arfgen} and {\it rmfgen} SAS tasks were used to create ancillary and response files.
The number of counts per channel were chosen depending on the source photon statistics.
For the sake of simplicity we discuss
data of the EPIC pn cameras only, while the MOS1 and MOS2 spectra have been checked for consistency.

In Figure~\ref{figure=lcurve} we show the XMM-\textit{Newton} light curves (binning time of 100 s)
between 0.3-2 keV (top panel), 2-10 keV (middle panel) and the hardness ratio (2-10 keV/0.3-2 keV)
(bottom panel) for each source. In Table~\ref{table=lc} we report the $\chi^{2}$/dof obtained by
fitting the light curves and hardness ratios with a constant; we have marked in bold face the 
$\chi^{2}$/dof with variability at less than 3 sigma confidence level. It is clear that, 
in both energy ranges, the short term light curves
show significant variability, in 50\% of the cases; the soft flux
is more variable than the 2-10 keV flux, as expected in NLSy1. 
The only exceptions are IRAS 15091-2107 and IGR J16185-5928, where the 2-10 keV flux is more variable than the soft flux.
The hardness ratios reflect some evidence of spectral variability, although this is more prominent
only in IGR J19378-0617 and indeed, this issue is investigated in details in a dedicated {\it Suzaku} observation
(Miniutti et al. in preparation). Given the variable nature of NLSy1 a time-resolved spectral analysis would be advisable, however
the XMM-\textit{Newton} and XRT exposures are too short for a statistically significant analysis,
therefore we extract the spectra from the total light curves.
 
With the aim of deriving the optical-to-X-ray spectral slope 
$\alpha_{ox}$~\footnote{$\alpha_{ox}$ $= -0.384log(f_{2 keV}/f_{2500 \AA})$, Tananbaum et al. (1979)}, 
optical U and UVW1 fluxes were extracted using the Optical Monitor (OM, Mason et al. 2001) data, already
available from the XMM-\textit{Newton} pipeline products. The background subtracted net count rates were converted
into fluxes by applying the OM count conversion factors. Finally a correction for Galactic reddening was applied 
using the E$_{B-V}$ values by Schlegel et al. (1998) and using 
the standard reddening correction curves by Cardelli et al. (1989).
In Table~\ref{table=opt_info} details of the OM data are reported.
 
 \begin{table*}
\caption{\bf XMM-\textit{Newton} Optical Monitor and {\it Swift}/UVOT Data}
\small{
\begin{center}
\begin{tabular}{lcccccc}
\hline
\hline
\multicolumn{1}{c}{Name} &
\multicolumn{1}{c}{Instrument} &
\multicolumn{1}{c}{Obs. Date} &
\multicolumn{1}{c}{Filter} &
\multicolumn{1}{c}{E$(B-V)$} &
\multicolumn{1}{c}{Optical Flux} &
\multicolumn{1}{c}{$\alpha_{ox}$} \\
\multicolumn{1}{c}{(1)} &
\multicolumn{1}{c}{(2)} &
\multicolumn{1}{c}{(3)} &
\multicolumn{1}{c}{(4)} &
\multicolumn{1}{c}{(5)} &
\multicolumn{1}{c}{(6)} &
\multicolumn{1}{c}{(7)} \\
\hline
\hline
1H 0323+342	    & XRT & 2006-07-05 & UVW1 & 0.213 	& 2.83$\pm$0.21 & 1.6\\
		    & XRT & 2006-07-09 & UVW1 & 	& 2.94$\pm$0.22 & 1.2\\
                    & XRT & 2007-11-04 & UVW1 & 	& 1.74$\pm$0.14 & 1.3\\
NGC4748		    & XRT & 2007-01-09 & UVW1 & 0.052	& 3.08$\pm$0.23 & 1.4 \\
Mrk 783		    & XRT & 2008-08-16 & UVW1 & 0.028	& 1.15$\pm$0.08 & 1.2\\
IGR J14552-5133     & XRT & 2006-12-27 & UVW1 & 0.662	& 4.69$\pm$0.48 & 1.4 \\
IRAS 15091-2107     & XMM & 2005-07-26 & UVW1 & 0.118	& 0.18$\pm$0.01 & 0.9	\\
IGR J16185-5928     & XMM & 2009-02-18 & U    & 0.322 	& 0.22$\pm$0.01 & 1.1	   \\  
IGR J19378-0617     & XMM & 2009-04-28 & U    & 0.293	& 2.72$\pm$0.03 & 1.1\\
ESO 399-IG 020	    & XRT & 2010-08-27 & UVW1 & 0.113	& 0.98$\pm$0.08 & 1.5\\
Swift J2127.4+5654  & XMM & 2009-11-11 & UVW1 & 1.298 	& 294.0$\pm$18.9& 1.8	\\
\hline
\end{tabular}
\end{center}
Notes: (1): Galaxy name; (2): Instrument; (3): Observation date;
(4): OM/UVOT filter; (5): (5): Optical flux in units of 10$^{-11}$ erg cm$^{-2}$ s$^{-1}$;
(6): Optical-to-X-ray slope $\alpha_{ox}$.} 
\label{table=opt_info}
\end{table*}

\subsection{The {\it Swift}/XRT data}

For five sources in our sample, we also have X-ray observations acquired with the XRT on board the \emph{Swift} satellite.
XRT data reduction was performed using the XRTDAS standard data pipeline package ({\sc xrtpipeline} v.
0.12.4), in order to produce screened event files. All data were extracted only
in the Photon Counting (PC) mode (Hill et al. 2004), adopting the standard grade filtering (0--12 for
PC) according to the XRT nomenclature. 
Events for spectral analysis were extracted within a circular region of radius 47$^{\prime \prime}$ (corresponding to 20 XRT pixels), 
centered on the source position,
which encloses about 90\% of the PSF at 1.5 keV (see Moretti et al. 2004).
The background was taken from various source-free regions close to the X-ray source of interest, using
circular regions with different radii in order to ensure an evenly sampled background. In all cases, the
spectra were extracted from the corresponding event files using the {\sc XSELECT} software and binned to 20 counts per bin
using {\sc grppha}, so that the $\chi^{2}$ statistic could be applied. We used
version v.011 of the response matrices and create individual ancillary response files \textit{arf} using
{\sc xrtmkarf v.0.5.6}.

1H 0323+342, NGC 4748, Mrk 783 and IGR J14552-5133 have multiple observations with XRT and, to increase the spectral
statistics, we fit the data sets together, tying the different parameters only in case of no or little 
variability. The strongest flux variation is observed in the blazar-like NLSy1 1H 0323+342,
where its flux has increased by a factor of $\sim$ 10 after two days. Unfortunately, the {\it INTEGRAL} observations were performed
between July and August 2004 and therefore, are not simultaneous with XRT data (see also Foschini et al. 2009,
for a presentation of the X-ray and hard X-ray behaviour of this source).
In NGC~4748, no spectral variation is found between the two $\sim$ 2 ks XRT observations,
therefore we tie the XRT photon indeces to the same value in the fitting procedure and report in Table~\ref{table=best}
the fluxes and cross-calibration constants.
In Mrk 783, the two XRT observations present both flux and spectral variability, while
neutral absorption is significantly required only in the first observation. 
Only little spectral and flux variation is found in IGR~J14552-5133.

We have estimated the 2500 \AA\ flux using data from the UVW1 filter of the UV/Optical 
Telescope (UVOT, Roming et al. 2005) onboard {\it Swift}.
From each segment data were co-added by the UVOT task {\it uvotimsum}. A circle of 3$^{\prime \prime}$ radius centered 
on the source has been used to extract source counts, taking into account the aperture correction. An annular background region centered on the source
is used, with an inner radius of 27.5$^{\prime \prime}$ and a width of 7.5$^{\prime \prime}$, except for IGR~J14552-5133
for which an off-source circle (with a radius of 15$^{\prime \prime}$) was used to avoid contamination from nearby sources. 
The data were analyzed with the UVOT software tool {\it uvotsource}, using the count rate to flux/magnitude conversion as in Poole et al. (2008). 
Magnitudes were corrected for Galactic reddening using the E$_{B-V}$ values by Schlegel et al. (1998) and using 
the standard reddening correction curves by Cardelli et al. (1989). In Table~\ref{table=opt_info} we report
the UVOT data log. Given the extreme variability between the three XRT observations of 1H 0323+342,
we report the three simultaneous UVOT/XRT $\alpha_{ox}$. 

\subsection{Broad-band spectral analysis}

\begin{table*}
\footnotesize{
\caption{\bf Combined XMM-\textit{Newton}/XRT and {\it INTEGRAL} continuum spectral analysis}
\label{table=best}
\begin{center}
\begin{tabular}{lccccccccr}
\hline
\hline
\multicolumn{1}{c}{Name} &
\multicolumn{1}{c}{$N_{\rm H, int}$} &
\multicolumn{1}{c}{Cvr} &
\multicolumn{1}{c}{$\Gamma_{0.3-100 keV}$} &
\multicolumn{1}{c}{C} &
\multicolumn{1}{c}{F$_{\rm 0.1-2}$} &
\multicolumn{1}{c}{F$_{\rm 2-10}$} &
\multicolumn{1}{c}{L$_{\rm 0.1-2}$} &
\multicolumn{1}{c}{L$_{\rm 2-10}$} &
\multicolumn{1}{c}{$\chi^{2}$/dof} \\
\multicolumn{1}{c}{(1)} &
\multicolumn{1}{c}{(2)} &
\multicolumn{1}{c}{(3)} &
\multicolumn{1}{c}{(4)} &
\multicolumn{1}{c}{(5)} &
\multicolumn{1}{c}{(6)} &
\multicolumn{1}{c}{(7)} &
\multicolumn{1}{c}{(8)} &
\multicolumn{1}{c}{(9)} &
\multicolumn{1}{c}{(10)}  \\
\hline
\hline
1H 0323+342 XRT1	      	&  $-$ & $-$ & 1.98$^{+0.16}_{-0.09}$ & 6.70$^{+0.82}_{-0.66}$  &  0.15 & 0.18 					& 43.06  &  43.14& 246/205 \\
1H 0323+342 XRT2	      	&  $-$ & $-$ & 2.05$^{+0.05}_{-0.05}$ & 3.50$^{+0.54}_{-0.45}$  &  0.77 & 1.11 					& 43.77  &  43.93& 246/205 \\
1H 0323+342 XRT3	      	&  $-$ & $-$ & 1.99$^{+0.13}_{-0.12}$ & 24.9$^{+21.2}_{-11.4}$  &  0.40 & 0.64 					& 43.49  &  43.69& 246/205 \\
NGC 4748 XRT1	        	&  $-$ & $-$ & 2.01$^{+0.13}_{-0.13}$ & 1.47$^{+0.20}_{-0.18}$  &  0.35 & 0.34 					& 42.19  &  42.18& 40/41 \\
NGC 4748 XRT2	        	&  $-$ & $-$ & 2.01$^{+0.13}_{-0.13}$ & $>$0.37                 &  0.52 & 0.50 					& 42.36  &  42.35& 40/41 \\
Mrk 783 XRT1		        &  0.15$^{+0.09}_{-0.08}$ & $-$ & 1.75$^{+0.23}_{-0.22}$ &  1.88$^{+0.64}_{-0.40}$ &  0.30 & 0.41		& 43.45  &  43.59& 81/80\\
Mrk 783 XRT2		        &  $>$ 0.11               & $-$ & 1.65$^{+0.17}_{-0.15}$ &  2.09$^{+2.09}_{-1.11}$ &  0.41 & 0.89		& 43.59  &  43.92& 81/80\\
IGR J14552-5133 XRT1   		&  $-$ & $-$ & 1.91$^{+0.10}_{-0.10}$ & 1.08$^{+0.15}_{-0.12}$ & 0.33 & 0.96 					& 42.24  &  42.71& 72/89 \\
IGR J14552-5133 XRT2   		&  $-$ & $-$ & 2.11$^{+0.10}_{-0.10}$ & 1.19$^{+0.48}_{-0.34}$ & 0.34 & 0.77 					& 42.26  &  42.61& 72/89 \\
IRAS 15091-2107         	&  0.14$^{+0.04}_{-0.04}$ & 0.75$^{+0.04}_{-0.04}$ & 1.75$^{+0.04}_{-0.03}$ & 1.54$^{+0.56}_{-0.50}$ &0.47& 0.90& 43.29  &  43.57& 431/480  \\
IGR J16185-5928         	&10.6$^{+16.0}_{-5.5}$ & 0.47$^{+0.19}_{-0.18}$  & 2.00$^{+0.09}_{-0.09}$ &6.23$^{+2.80}_{-2.68}$ & 0.13 & 0.27 & 42.52  &  42.84& 185/212	 \\
IGR J16385-2057 XMM1        	&  $-$ & $-$ & 1.50$^{+0.05}_{-0.07}$ & 0.60$^{+0.21}_{-0.18}$ & 0.40 & 0.70					& 42.78  &  43.02& 593/606 \\  
IGR J16385-2057 XMM2        	&  $-$ & $-$ & 1.84$^{+0.06}_{-0.06}$ & 1.64$^{+0.51}_{-0.46}$ & 0.32 & 0.53					& 42.68  &  42.90& 595/551 \\  
IGR J19378-0617         	& 32.0$^{+10.5}_{-8.0}$ & 0.47$^{+0.05}_{-0.05}$ & 2.58$^{+0.02}_{-0.02}$ & 1.37$^{+0.40}_{-0.35}$&3.4 & 2.7  	& 42.90  &  42.80& 487/465 \\
ESO 399-IG 020         		& $-$ & $-$ & 1.69$^{+0.17}_{-0.17}$ & $>$0.80 & 0.07 & 0.13  							& 41.95  &  42.22& 19/22\\
Swift J2127.4+5654     		& $-$ & $-$ & 1.94$^{+0.07}_{-0.08}$ &0.84$^{+0.21}_{-0.17}$ & 1.0 & 4.2  					& 42.61  &  43.23& 1337/1265$^{*}$ \\
\hline				
\end{tabular}
\end{center}
Note: (1): Galaxy name; (2): Intrinsic fully or partially covering column density in units of 10$^{22}$ cm$^{-2}$;
(3): Covering factor; (4) Power-law photon index; (5): Value of the XMM-\textit{Newton}/{\it INTEGRAL} and XRT/{\it INTEGRAL} cross-calibration constant; 
(6-7): Model unabsorbed fluxes in the 0.1-2 keV and 2-10 keV bands in units 
of 10$^{-11}$ erg cm$^{-2}$ s$^{-1}$; (8-9): Logarithm  of the unabsorbed luminosities in the 0.1-2 keV and 2-10 keV bands;
(10): Chi-squared and degrees of freedom.
$^{*}$ Best fit obtain with a {\it pexrav} model and the addition of an extra gaussian component at 5.8 keV (see Section~\ref{Feprox}).}
\end{table*}

\begin{table*}
\footnotesize{
\caption{\bf Fe lines and soft excess spectral analysis.}
\label{table=fe}
\begin{center}
\begin{tabular}{llllcccccr}
\hline
\hline
\multicolumn{1}{c}{Name} &
\multicolumn{1}{c}{$E_\alpha$} &
\multicolumn{1}{c}{$\sigma$} &
\multicolumn{1}{c}{$EW_\alpha$} &
\multicolumn{1}{c}{$E_\alpha$} &
\multicolumn{1}{c}{$\sigma$} &
\multicolumn{1}{c}{$EW_\alpha$} &
\multicolumn{1}{c}{$kT$} &
\multicolumn{1}{c}{$kT_{in}$} &
\multicolumn{1}{c}{$\chi^{2}$/dof} \\
\multicolumn{1}{c}{(1)} &
\multicolumn{1}{c}{(2)} &
\multicolumn{1}{c}{(3)} &
\multicolumn{1}{c}{(4)} &
\multicolumn{1}{c}{(5)} &
\multicolumn{1}{c}{(6)} &
\multicolumn{1}{c}{(7)} &
\multicolumn{1}{c}{(8)} &
\multicolumn{1}{c}{(9)} &
\multicolumn{1}{c}{(10)} \\
\hline
\hline
IRAS 15091-2107         	& 6.31$^{+0.14}_{-0.10}$ & 0.19$^{+0.26}_{-0.07}$ & 170$^{+150}_{-60}$ &  $-$ & $-$ & $-$ &  $-$ & $-$ & 431/480 \\
IGR J16385-2057 XMM1        	& 6.37$^{+0.07}_{-0.18}$ & 0.01$^{fix}$ & 71$^{+47}_{-47}$ & $-$ & $-$ & $-$  & 0.60$^{+0.09}_{-0.13}$ & 0.48$^{+0.03}_{-0.02}$ & 593/606  \\  
IGR J16385-2057 XMM2        	& 6.60$^{+0.06}_{-0.06}$ & 0.01$^{fix}$ & 75$^{+38}_{-38}$ &6.97$^{+0.04}_{-0.05}$ &0.01$^{fix}$ & 90$^{+33}_{-26}$ & 0.13$^{+0.03}_{-0.03}$ & 0.33$^{+0.03}_{-0.03}$ & 595/551 \\
IGR J19378-0617         	&  $-$&   $-$ & $-$ & $-$&   $-$ & $-$ & 0.22$^{+0.01}_{-0.01}$&$-$&  487/465	 \\
Swift J2127.4+5654     		& 6.40$^{+0.04}_{-0.04}$ & 0.01$^{fix}$ & 46$^{+14}_{-14}$&5.80$^{+0.04}_{-0.04}$ & 0.01$^{fix}$ & 38$^{+22}_{-20}$ &0.22$^{+0.01}_{-0.01}$ & $-$ & 1337/1265 \\
\hline				
\end{tabular}
\end{center}
Note: (1): Galaxy name; (2): Energy of the Fe line in keV; (3): Line width in keV;
(4): Equivalent width of the Fe line in eV; (5): Energy of a second gaussian component in keV; (6): Line width in keV;
(7): Equivalent width in eV; (8):  Plasma temperature in keV; (9): {\it DISKBB} temperature at inner disk radius (keV);
(10): Chi-squared and degrees of freedom.}
\end{table*}

Since the XMM-\textit{Newton}/XRT and {\it INTEGRAL} observations 
are not simultaneous, a cross-calibration constant $C$ is left free to vary
to take into account possible cross-calibration mismatches between
the two instruments but more importantly in these objects, to consider variability effects.
The cross-calibration constants between {\it INTEGRAL} and XMM-\textit{Newton}, XRT and {\it Chandra}
have been typically found narrowly distributed around unity using various
source typologies (e.g. Molina et al, 2009, De Rosa et al. 2008; Masetti et al. 2007; Panessa et al. 2008), 
thus suggesting that significant deviation from this value is confidently 
attributed to source flux variations.

With the aim of characterizing broadly the spectral
properties of the sample, we have added
extra components to a baseline model composed of a simple power-law with absorption
fixed at the Galactic value, to model spectral residuals. 

Both an emission line spectrum from hot diffuse gas ({\it MEKAL} in Xspec) and/or a disk black body component ({\it DISKBB} in Xspec)
has been tested to model the soft X-ray emission; in more complex spectra,
absorption edges were added with energies ranging from 0.7 to 2 keV, as expected from
a "warm" absorber (Reynolds 1997).

\begin{figure*}
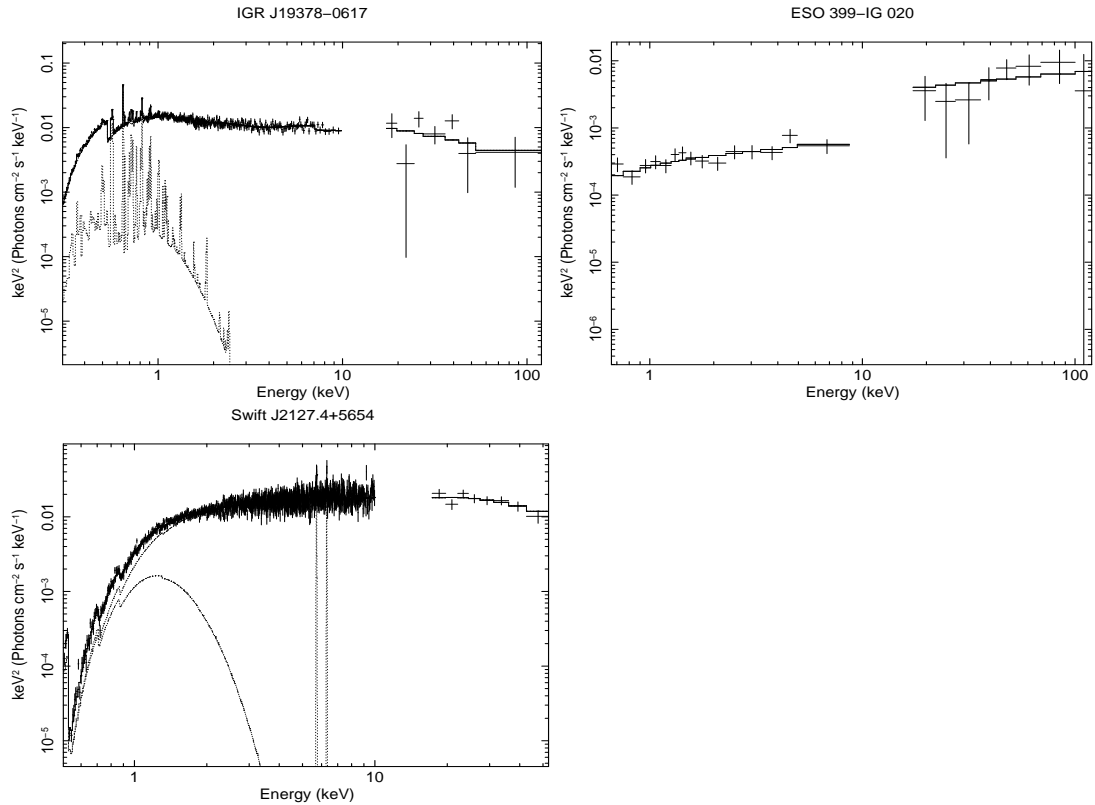

\begin{center}
\parbox{16cm}{
\includegraphics[width=0.3\textwidth,height=0.3\textheight,angle=-90]{IGRJ19378_spec.ps}
\includegraphics[width=0.3\textwidth,height=0.3\textheight,angle=-90]{ESO399-20_spec.ps}}
\parbox{16cm}{
\includegraphics[width=0.3\textwidth,height=0.3\textheight,angle=-90]{Swift2127_spec_pexrav.ps}}
\caption{Broad-band XMM-\textit{Newton}/XRT and {\it INTEGRAL} best--fit spectra.}
\label{figure=bestfit2}
\end{center}
\end{figure*}

Residuals around 6-7 keV have been modeled with one 
or two gaussian components representing the Fe line neutral/ionized
transitions. Whenever necessary, the curvature of the continuum was modeled with
a partial covering absorption. The cut-off power-law and {\it pexrav} models were tested on each source.
However, the best-fit model for all the sources in the sample is achieved by using 
a simple power-law model, except for Swift~J2127.4+5654 for which a {\it pexrav} model better
represents data.  
The spectral fitting results are reported in Table~\ref{table=best} and broad-band spectra
are shown in Figure~\ref{figure=bestfit1} and \ref{figure=bestfit2}.
Spectral best fitting was performed using XSPEC v12.5.0 (Arnaud 1996).
The quoted errors on the model parameters correspond to a 90\% confidence level for
one interesting parameter. 

The broad-band spectrum for three sources of our sample (i.e., NGC~4051, Mrk~766 and NGC~5506) has previously
been studied in the literature (Terashima et al. 2009, Turner et al. 2007 and Guainazzi et al. 2010, respectively).
For the purpose of a general broad-band statistical analysis of the total sample, we have used the 
best-fit spectral parameters reported in literature for these three sources,
using the averaged spectral values and fluxes.
Similarly, for the sources analyzed in this work having multiple observations, we have used
the averaged parameters in our plots.

\begin{figure*}
\begin{center}
\parbox{16cm}{
\includegraphics[width=0.3\textwidth,height=0.3\textheight,angle=-90]{1H_best.ps}
\includegraphics[width=0.3\textwidth,height=0.3\textheight,angle=-90]{4748_spec.ps}}
\parbox{16cm}{
\includegraphics[width=0.3\textwidth,height=0.3\textheight,angle=-90]{Mrk783_spec.ps}
\includegraphics[width=0.3\textwidth,height=0.3\textheight,angle=-90]{14552_spec.ps}}
\parbox{16cm}{
\includegraphics[width=0.3\textwidth,height=0.3\textheight,angle=-90]{IRAS15091_spec.ps}
\includegraphics[width=0.3\textwidth,height=0.3\textheight,angle=-90]{IGRJ16185_spec.ps}}
\parbox{16cm}{
\includegraphics[width=0.3\textwidth,height=0.3\textheight,angle=-90]{IGRJ16385xmm1_best.ps}
\includegraphics[width=0.3\textwidth,height=0.3\textheight,angle=-90]{IGRJ16385xmm2_spec.ps}}
\caption{Broad-band XMM-\textit{Newton}/XRT and {\it INTEGRAL} best--fit spectra.}
\label{figure=bestfit1}
\end{center}
\end{figure*}

\section{Results}

\subsection{Continuum slope and high energy cut-off}

\begin{figure*}
\begin{center}
\parbox{16cm}{
\includegraphics[width=0.47\textwidth,height=0.3\textheight,angle=0]{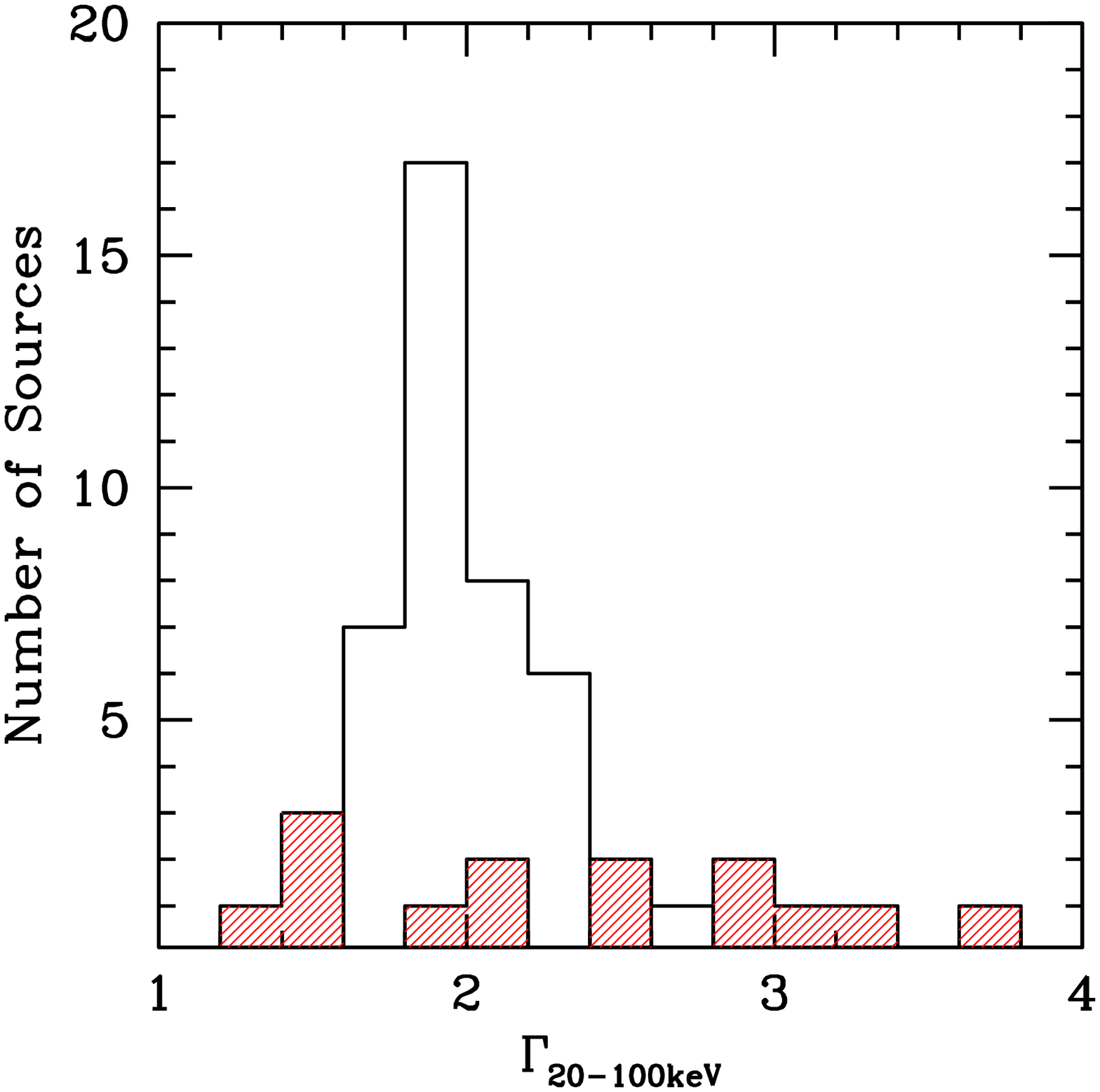}
\includegraphics[width=0.47\textwidth,height=0.3\textheight,angle=0]{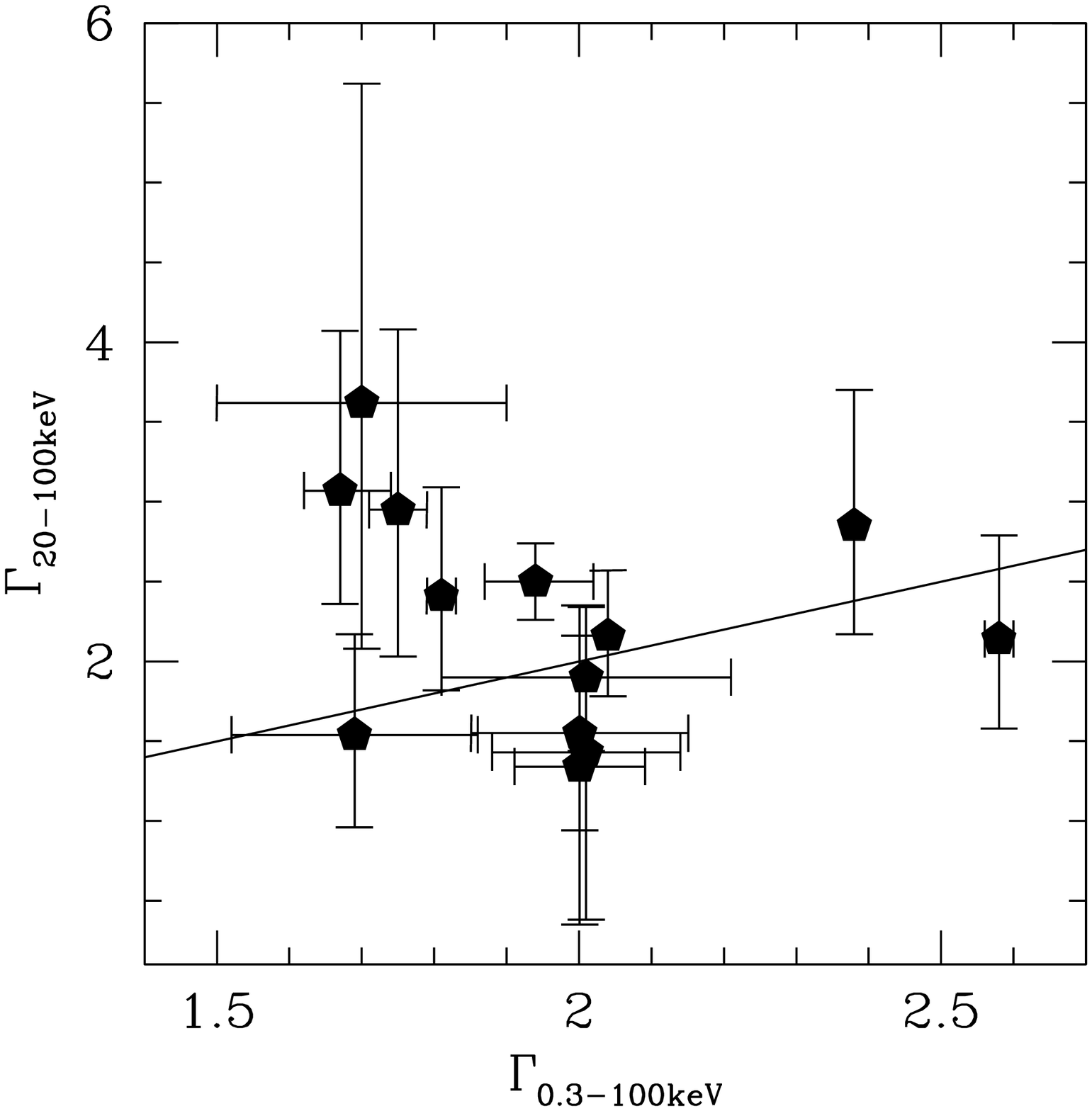}}
\caption{Left panel: Distribution of the 20-100 keV photon index for our sample of NLSy1 (shaded histogram)
and a sample of hard X-ray selected BLSy1 (Molina et al. in preparation). Right panel: 20-100 keV photon index versus
the 0.3-100 keV broad-band photon index in NLSy1.}
\label{figure=foton}
\end{center}
\end{figure*}

In Figure~\ref{figure=foton} (left panel), the hard photon index distribution in NLSy1 sample
(shaded histogram) is compared to the {\it INTEGRAL} BLSy1 one (IBIS fourth catalogue, Molina et al. in preparation).
The hard photon index is broadly distributed in our sample ranging from $\sim$ 1.3 to very steep values such as 
$\sim$ 3.6, with a mean value $<$$\Gamma_{20-100 keV}$$>$ $=$ 2.3$\pm$0.7~\footnote{We use the arithmetic mean 
and adopt its standard deviation as a measure of each parameter spread, given the extreme non-Gaussian distribution of the data points.} 
(see Table~\ref{table=hard}). This value is 
consistent within errors with the one previously found in a smaller sample of five {\it INTEGRAL} NLSy1 ($<$$\Gamma_{20-100 keV}$$>$ $=$ 2.6$\pm$0.3,
Malizia et al. 2008). 
The BLSy1 mean hard photon index ($<$$\Gamma_{20-100 keV}$$>$ $=$ 2.0$\pm$0.2) is only slightly flatter and consistent 
with the NLSy1 $<$$\Gamma_{20-100 keV}$$>$, as indeed a Kolmogorov-Smirnov (K-S) test (probability $P = 0.030$ of a random result) indicates 
that the distributions are not significantly different. This is also in agreement with estimates from type 1 Seyfert spectra from {\it Swift}/BAT
($<$$\Gamma_{20-100 keV}$$>$ $=$ 2.23$\pm$0.11, Ajello et al. 2008). From this analysis, the average photon index of NLSy1 at hard X-rays appears
not to be steeper than in BLSy1, however, we should consider that faint steep spectrum sources may be missed by hard X-ray surveys.

By fitting the 0.3-100 keV broad-band spectrum (as in Table~\ref{table=best} and including data from literature
for NGC~4051, Mrk~783 and NGC~5506), the resulting
photon index varies from $\sim$ 1.5 to $\sim$ 2.6 with a mean of $<$$\Gamma_{0.3-100 keV}$$>$ $=$ 2.0$\pm$0.3,
consistent with the typical values found for this class of sources (e.g., Leighly 1999).  
We should note that the 0.3-100 keV spectral fit parameters are more representative of the
spectra below 10 keV, for the larger statistics, especially when XMM-\textit{Newton} data are used.
In a complete hard X-ray selected sample of BLSy1, Molina et al. (2009) found a flatter mean value $<$$\Gamma_{20-100 keV}$$>$ $\sim$ 1.7
($\sigma$ $=$ 0.2), confirming the evidence that NLSy1 tend to have steeper photon indeces, as also confirmed in Bianchi et al. (2009a),
for an X-ray selected sample. Therefore, no clear separation between NLSy1 and BLSy1 average hard X-ray photon index is found
as instead observed for the broad-band photon index. 

In Figure~\ref{figure=foton} (right panel) we plot the hard X-ray 
versus the 0.3-100 keV broad band X-ray photon index. The one-to-one regression line is also drawn.
The steeper hard X-ray photon indeces are also clear from this plot suggesting the possible presence
of a high energy cut-off. However, with the present data we were able to constrain
this parameter only in Swift~J2127.4+5654 (E$_{cut-off}$ $=$ 49$^{+49}_{-17}$ keV), 
in agreement with our previous {\it INTEGRAL} estimate (Malizia et al. 2008) and with 
the {\it Suzaku} PIN measurement (Miniutti et al. 2009).
A very steep hard X-ray photon index is measured in Mrk 766 ($\Gamma$ $\sim$ 2.9) and indeed,
a spectral decrease is evident at $\sim$ 50 keV from the {\it Suzaku} PIN data (Turner et al. 2007).
In Mrk~783,  IRAS 15091-2107 and in IGR~J16385-2057 the hard X-ray photon index is much steeper than the XRT and XMM-\textit{Newton} one,
however the cut-off energy measured from the broad-band analysis would be physically unexpected (E$_{cut-off}$ $\sim$ 20 keV).
Instead, the XMM-\textit{Newton}/XRT and {\it INTEGRAL} mismatch is likely due to spectral and/or flux variability. 
Finally, IGR~J16426+6536 shows a steep hard X-ray photon index,
however the lack of data below 10 keV does not allow a constraint on the broad-band spectrum.
 
\subsection{Fe K emission lines and reprocessing features}\label{Feprox}

Fe emission lines are ubiquitously found in Seyfert spectra (e.g., Nandra et al. 1997).
Indeed, the origin of both the narrow and broad component is still controversial and
NLSy1 often do show complex, variable and multiple components, as in the case of
NGC~4051, Mrk766 and NGC~5506 (for a comprehensive report on the Fe line properties see 
Terashima et al. 2009, Turner et al. 2007 and Guainazzi et al. 2010 respectively).

A fluorescent emission line around 6-7 keV is detected in three out of four NLSy1 observed by XMM-\textit{Newton}
(see Table~\ref{table=fe}), this fraction increases to six out of seven if we consider the line detections in NGC~4051, Mrk~766 and NGC~5506.
We must consider that no data at these energies are available in the IGR~J16185-5928 XMM-\textit{Newton} spectrum since it is dominated 
by the background, due to high flaring activity in a short exposure observation. The Fe line is not significantly
required in the IGR J19378-0617 spectral fitting, although a weak feature is present.
In IRAS 15091-2107 the line profile is consistent with a moderately broad Gaussian component,
while between the two XMM-\textit{Newton} observations of IGR J16385-2057 the line has varied its energy; also, marginal evidence
for a narrow line at $\sim$ 7 keV (EW $=$ 60$^{+50}_{-40}$ eV) in both observations suggest the presence of variable ionized Fe lines
in this source. 
A broad relativistic Fe line has been detected in a {\it Suzaku} observation of Swift~J2127.4+5654 (Miniutti et al. 2009).
Time-resolved spectroscopy performed on a small portion of the light curve when the source is at an higher flux state reveals the presence of two
narrow Gaussian line at 5.83$\pm$0.12 keV and at 6.57$\pm$0.10 keV. We confirm this evidence 
and significantly detect an emission line at 5.8 keV.
A deeper investigation on the line properties of this source is ongoing using a longer ($\sim$ 100 ks) 
XMM-\textit{Newton} observation (Miniutti et al. in preparation). 

Overall, the modeling of the broad-band X-ray spectrum with a 
cut-off power-law continuum reflected from neutral material ({\it pexrav}) does not provides a significant
increase of the fit quality. In
the case of IGR J16185-5928, the XMM-\textit{Newton} and {\it INTEGRAL} flux difference
results in an extremely high reflection value $R$=14.9$^{+9.3}_{-7.4}$ 
(with C$_{pn/IBIS}$=1.4$^{+0.6}_{-0.4}$ instead of $\sim$ 6.2 when modeling with a power-law);
this is probably to be ascribed to variability as shown from the {\it Suzaku} longer observation, where
a more moderate reflection fraction is measured (R $=$ 0.9 $\pm$ 0.2, Miniutti et al. in preparation).
The 20-100 keV {\it Suzaku} flux of 1.5 $\times$ 10$^{-11}$ erg cm$^{-2}$ s$^{-1}$ is consistent
with the average {\it INTEGRAL} flux of 1.7 $\times$ 10$^{-11}$ erg cm$^{-2}$ s$^{-1}$, while 
during the XMM-\textit{Newton} observation the 2-10 keV flux was a factor of $\sim$ 3 lower than that of {\it Suzaku}.  
The reflection fraction is well constrained in Swift~J2127.4+5654 (R=1.1$^{+0.8}_{-0.7}$)
in agreement with the {\it Suzaku} value (R=1.0$^{+0.5}_{-0.4}$, Miniutti et al. 2009).

\subsection{Soft X-ray components and absorption}

The soft X-ray emission is one of the most peculiar characteristics in NLSy1
for its complex shape and variability. In this work, only phenomenological
models are used to characterize the soft emission, such as a disk
black body model and an emission line spectrum from hot diffuse gas.
It has been noticed that the use of a thermal disk model in AGN results in a very narrow range of disk
temperatures independently on the range of luminosities and black hole masses spanned (Ponti et al. 2006).
Different models involving atomic physics processes could certainly be more adequate to describe
the soft X-ray emission (e.g., Gierli{\'n}ski \& Done 2004), however this is beyond the scope of this work where
a simple representation of the average properties is instead provided.

In our sample, a {\it MEKAL} model well represents the soft emission in all sources except for IRAS~15091-2107 and IGR~J16185-5928
for which no evidence of soft X-ray emission is found. The temperatures are in the range between 0.1 to 0.6 keV.
The addition of a black body model is also required
in the spectral fitting of IGR~J16385-2057, where
the temperature varies between the two observations as well as a larger variation is seen in the temperature
of the hot plasma, suggestive of spectral variability also at soft energies.
The presence of absorption edges due to O VII (0.737 keV) and O VIII (0.871 keV) is 
not required in the sources spectral fitting. 

The most complex and prominent soft spectrum is shown by
IGR~J19378-0617 where two absorption edges are needed to model the spectral features
at 1.23$\pm$0.03 keV ($\tau$=0.13$\pm$0.02) and 1.57$\pm$0.03 keV ($\tau$=0.16$\pm$0.03); 
such features have been previously detected in AGN and interpreted in terms of highly blue-shifted absorption material,
resonance absorption by Fe L (see Leighly 1999), as expected from partially ionized absorption by a warm
absorber and/or reflection off the accretion disc.

The presence of strong, partial and/or stratified absorption has also been
found in NGC 4051, Mrk 766 and NGC 5506 and is one of the competing models applied in objects 
with complex spectra.
Intrinsic absorption was measured in four sources of the sample, in three objects it partially covers 
the continuum and in IGR~J19378-0617 this value is significantly high ($\sim$3 $\times$ 10$^{23}$ cm$^{-2}$),
similar to the amount of obscuration measured in NGC 5506.

To broadly characterize the prominence of the soft X-ray emission with respect to the total X-ray spectrum,
we have adopted the soft X-ray luminosity (0.1-2 keV) and the hard X-ray luminosity (2-10 keV) 
ratio as in Bianchi et al. (2009a), based on the fact that luminosities are only slightly affected by the choice of the model. 
Bianchi et al. (2009a) found that NLSy1 have luminosity ratios higher than unity and are distributed at higher values
with respect to BLSy1, showing a soft X-ray emission dominance in NLSy1.
This is not confirmed in our sample, where only in IGR~J19378-0617 and NGC~4748 (see Table~\ref{table=best}) 
the luminosity ratio is only slightly higher than 1, suggesting that the hard X-ray selection of the sample
may reduce the bias introduced by the ROSAT soft X-ray selection of NLSy1.

\begin{figure*}
\begin{center}
\includegraphics[width=0.47\textwidth,height=0.3\textheight,angle=0]{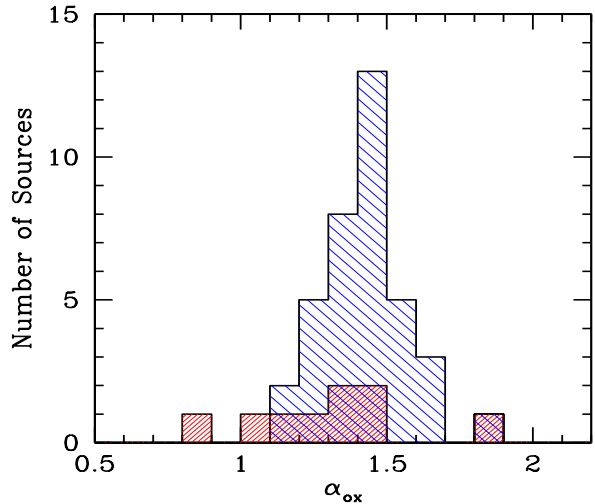}
\caption{Distribution of the optical-to-X-ray spectral slope $\alpha_{ox}$ in our sample (right-had shaded histrogram)
and in Grupe et al. (2010) NLSy1 sample (left-hand shaded histogram).}
\label{figure=aox}
\end{center}
\end{figure*}

\begin{figure*}
\begin{center}
\parbox{16cm}{
\includegraphics[width=0.47\textwidth,height=0.3\textheight,angle=0]{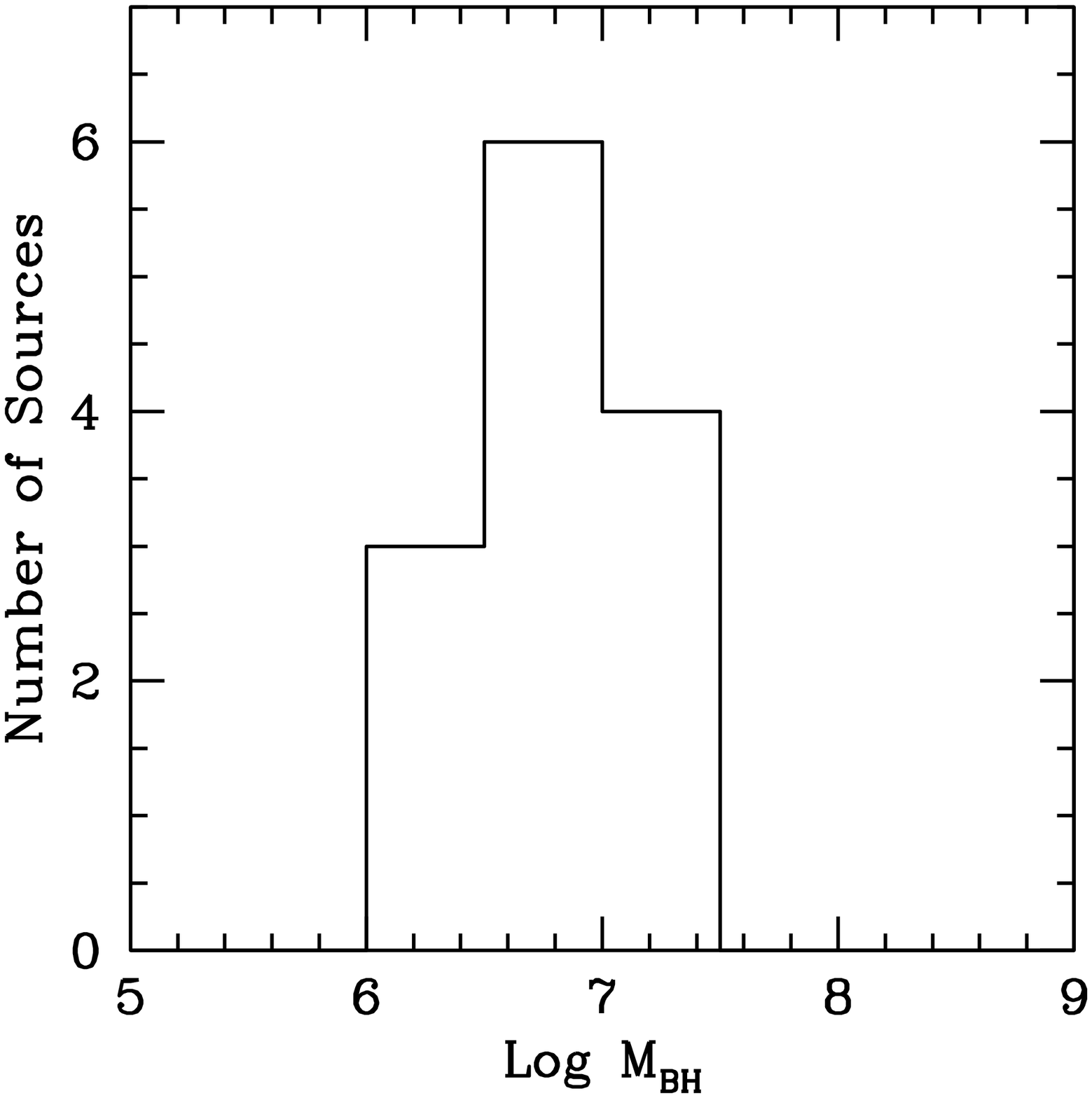}
\includegraphics[width=0.47\textwidth,height=0.3\textheight,angle=0]{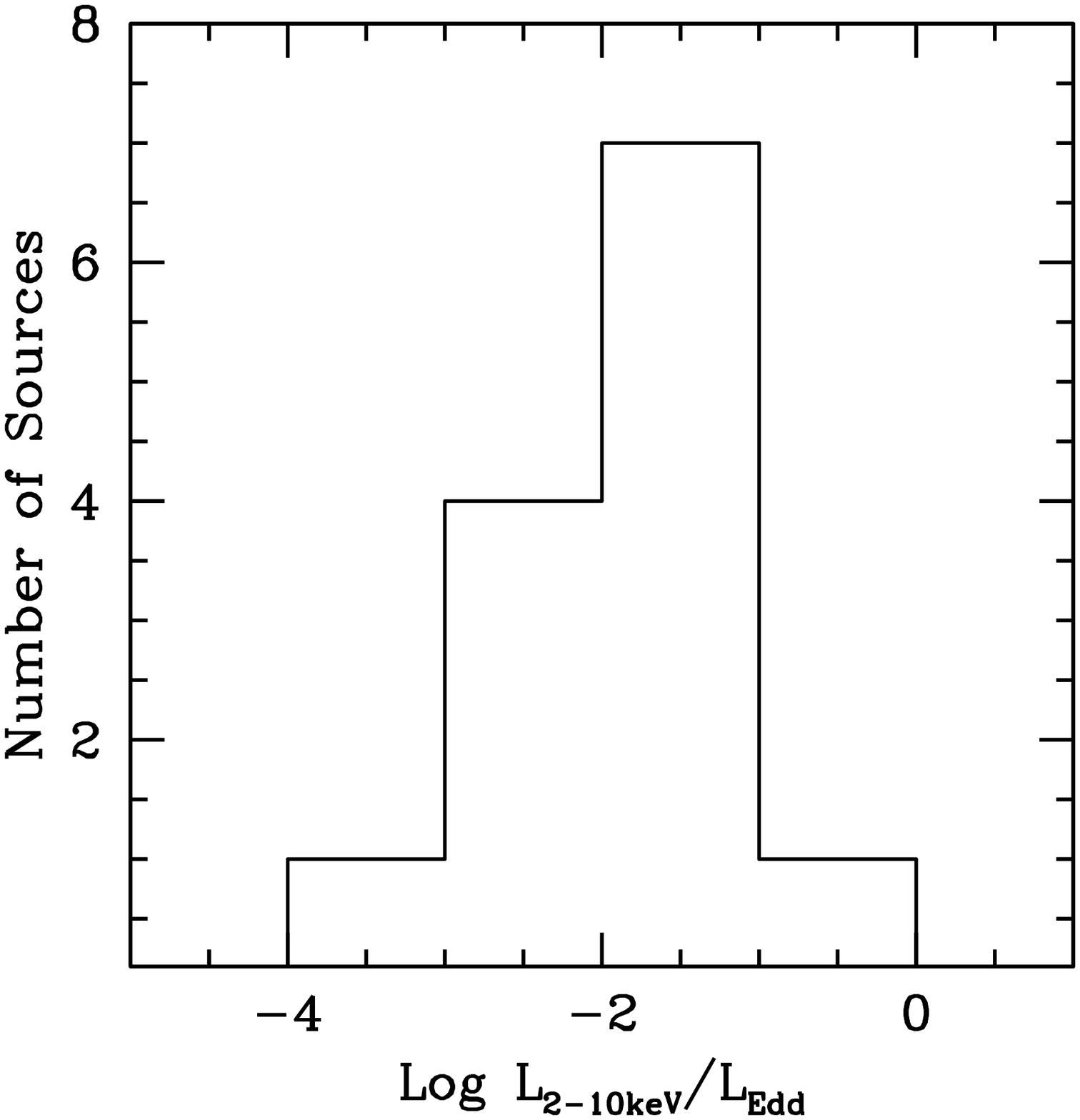}}
\caption{Left panel: Log black hole mass distribution in NLSy1. Right panel: Log L$_{2-10 keV}$/L$_{Edd}$ ratio
distribution.}
\label{figure=mass}
\end{center}
\end{figure*}

\subsection{Hard X-ray variability}

In Bird et al. (2010), sources variability has been searched for at revolution level, at revolution sequence (sequence of consecutive similar
pointings) and then also by optimizing the source detection time scale; this is the so-called "bursticity" analysis.  
In fact, light curve for each source has been scanned with variable time window to obtain the best source significance 
in the 18-60 keV band, then both duration and time-interval of the best detection period have been recorded. 
The ratio of the maximum significance on any
time-scale to the significance  for the whole data set define the "bursticity" so that a value of 1 is associated to 
persistent sources for which all data set can be used to increase its significance while a values greater than 1 indicates
variability and hence part of the data can be neglected to enhance the significance. Of course the higher is the bursticity values,
the strongest is the source variability. 
Only one source in our list, IGR~J16426+6536, has increased its flux by a factor of $\sim$ 3 on less than one day time scale,
the rest of the sample does not show burst-like events. It must be noted
that the source significance is often at the sensitivity limit not allowing a proper characterization
of the long term variability properties within their {\it INTEGRAL} light curves. 
NGC~4051 is the only source in our sample which belongs to the list of variable {\it INTEGRAL} sources
edited by Telezhinsky et al. (2010).

A comparison between the fourth IBIS catalogue and BAT 54 months shows that fluxes in the 20-100 keV energy range  
are almost consistent, with a maximum difference by a factor of 3, as in the case of 1H 0323+342.
The large variability in 1H 0323+342 has already been discussed (Foschini et al. 2009)
and expected given the blazar-like nature of this source. 
We should bear in mind that the IBIS fluxes are averaged in the period between 2002 and 2008, while BAT fluxes from 2004 till 2009,
therefore data partially overlaps. However, we notice that BAT fluxes tend to be systematically 
lower with respect to IBIS fluxes (except for NGC~5506). In a systematic comparison between IBIS and BAT data performed on 
a complete sample of {\it INTEGRAL} sources, it has been found that indeed there is a tendency for BAT fluxes to be
lower with respect to the IBIS fluxes, especially for dimmer sources (as NLSy1 are): this can be interpreted as a sign of some
systematics between the two instruments (Molina et al. in preparation).

On shorter time-scales, i.e., within a single {\it Suzaku}/PIN observation taken in 2005, NGC~4051 shows 
only moderate variability in the 15-40 keV light curve, from 2.4 to 2.9 $\times$ 10$^{-11}$ ergs cm$^{-2}$ s$^{-1}$
in the low and high flux states, respectively (Terashima et al. 2009); this values are consistent with the 
average {\it INTEGRAL} flux. The hard X-ray flux has increased during a second {\it Suzaku} observation in 2008, where 
a strong excess over a power law at energies 
above 20 keV is found, some fraction of which also appears to vary with the power-law continuum
(Miller et al. 2010). Mrk~766 was observed twice above 10 keV by the PDS on board {\it BeppoSAX} with a flux of 3 $\times$ 10$^{-11}$ erg cm$^{-2}$ s$^{-1}$
showing evidence of flux variability (Landi et al. 2005).
The 12-40 keV {\it Suzaku} light curve of Mrk~766 (Turner et al. 2007) is
consistent with a constant flux at around 0.01 counts s$^{-1}$, which translates into a 
12-40 keV flux of 4 $\times$ 10$^{-12}$ erg cm$^{-2}$ s$^{-1}$, 40\% lower than the {\it INTEGRAL} and BAT average flux
in that energy range.
The 20-100 keV {\it Suzaku} flux of IGR~J16185-5928 is 1.5 $\times$ 10$^{-11}$ erg cm$^{-2}$ s$^{-1}$ (Miniutti et al. in preparation), consistent 
with the average {\it INTEGRAL} value. Also for Swift~J2127.4+5654, the IBIS average flux is perfectly consistent 
with the {\it Suzaku} PIN flux (F$_{20-100  keV}$ = 4.2  $\times$ 10$^{-11}$ erg cm$^{-2}$ s$^{-1}$).
NGC~5506 is known to be one of the most variable AGN in the X-ray domain on short-time scales (e.g., $\sim$ 300 s, Dewangan \& Griffiths, 2005),
while on long-time scale a systematic variation on its spectral slope is found (Guainazzi et al. 2010). At energies above 20 keV,
a monitoring campaign with HEXIT (Manchanda 2006) reveals a variation by a factor of 10 in flux and also
spectral variability. The 15-136 keV {\it BeppoSAX} flux of 17 $\times$ 10$^{-11}$ ergs cm$^{-2}$ s$^{-1}$ (Deluit 2004)
is in agreement with the average {\it INTEGRAL} value.
Summarizing, in those sources with multiple targeted observations, variability is often found (three out of five sources in our sample),
suggesting that variability is a common trait in NLSy1 not only in X-rays but also in the hard X-rays domain.

\section{Discussion}

\subsection{Fraction of hard X-ray NLSy1}

In the fourth IBIS catalogue, 723 hard X-ray Galactic and extra-Galactic sources are listed,
nearly 30\% of these sources had an "unknown" classification/identification at the time of the publication of the catalogue.
The continuous follow-up campaigns for the identification of the new hard X-ray sources is now reducing this fraction.
As by August 2010 (Masetti et al. 2010), nearly 35\% of the sources in the catalogue have been identified as AGN (258), 
with 207 objects classified as Seyfert galaxies (98 as type 1
and 109 as type 2). Therefore, in the {\it INTEGRAL} hard X-ray sky, NLSy1 constitute $\sim$ 14\% of the type 1 Seyfert population and $\sim$ 5\% 
of the AGN population.
However, it should be considered that the hard X-ray sky is not uniformly covered by the {\it INTEGRAL} exposures: 70\% of the sky around
the Galactic plane region is covered to better than 1 mCrab sensitivity, compared to the 5 mCrab level of 90\% of the extragalactic regions (Bird et al. 2010).
This is likely introducing a bias against the number of AGN detections, which is indeed increasing during the latest years when the survey
is covering more extragalactic fields. The 
AGN (and NLSy1) number is therefore to be intended as a lower limit. 
On the other hand, extragalactic sources in the {\it Swift}/BAT sky largely outnumber the galactic sources, because of the uniform sky coverage (e.g., Cusumano et al. 2010, Tueller et al. 2010). 
In the Palermo {\it Swift}/BAT catalogue (54 months), 735 AGN are listed, of which 472 are Seyfert galaxies. 
The lack of a fine Seyfert classification does not allow a proper determination of the NLSy1 fraction. 
To our knowledge, 14 among the 307 type 1 Seyfert galaxies reported are indeed NLSy1, 
reducing their fraction to $\geq$ 5\% among type 1 AGN and only to $\geq$ 2\% among the total AGN population.
This low fraction could be possibly ascribed to the BAT lower exposure observation on each target or sky area. Moreover, 
the better IBIS sensitivity limit compared to the BAT instrument allows the detection of faint hard X-ray AGN such as NLSy1.
If we consider the complete hard X-ray sample in the 20-40 keV band by Malizia et al. (2009), the fraction
of NLSy1 is $\sim$ 11\% among type 1 Seyfert galaxies and $\sim$ 6\% among all AGN in the sample. This is probably
the more realistic estimate of the NLSy1 fraction in the hard X-ray sky.

An unbiased estimate of the fraction of NLSy1 among AGN is difficult to obtain, since this number strongly depends on the
sample selection and the energy range used. During the ROSAT soft X-ray survey a large number of NLSy1 have been
discovered; Grupe (2004) estimated a fraction of $\sim$ 46\% of NLSy1 while Hasinger et al. (2000) only $\sim$ 1\%.
An optically selected sample provides a NLSy1 fraction of $\sim$ 15\% among the type 1 AGN population  (Zhou et al. 2006, Williams et al. 2002),
in agreement with the hard X-ray fraction (5-15\%) which suggests that hard X-rays are efficient in selecting NLSy1.

\subsection{Black Hole mass, Eddington ratio and radio-loudness}

The peculiar properties of NLSy1 have been commonly associated to their small black hole masses and relative high Eddington ratios.
Black hole masses are usually estimated through mass-radius relations (see Table~\ref{sam}), the latter derived from the optical continuum luminosity (Kaspi et al. 2000).
However it must be kept in mind that large uncertainties affect these estimates. 
The distribution of the black hole masses in our sample (Figure~\ref{figure=mass}, left panel) peaks at $\sim$ 10$^{7}$ M$_{\odot}$, confirming the
findings that NLSy1 host preferentially small black holes and possibly suggesting that the hard X-ray origin of the sample
is not introducing any particular selection effect in the black hole mass range. This result is similar to that found in an X-ray selected sample 
where Bianchi et al. (2009b) have shown that NLSy1 have the smallest BH masses when compared to BLSy1.
The 2-10 keV luminosity and Eddington luminosity ratios (Figure~\ref{figure=mass}, right panel)
also point to highly accreting systems, although hard X-ray NLSy1 
seem to occupy the lower Eddington ratios tail of the NLSy1 population distribution (see Bianchi et al. 2009b).
A correlation between the photon index and the Eddington ratio has been found
by several authors (Grupe et al. 2010, Risaliti et al. 2009, Shemmer et al. 2008). 
Hard X-ray NLSy1 may occupy the "flat-$\Gamma$/low Eddington ratio" part of this correlation
with respect to the soft X-ray selected NLSy1 which may be found in the "steep-$\Gamma$/high Eddington ratio" part.

Grupe et al. (2010) have also shown that for a sample of X-ray selected NLS1y and BLSy1 the distribution
of the reddening corrected optical-to-X-ray spectral slopes $\alpha_{ox}$ was very similar. To check for a possible
difference introduced by the hard X-ray selection, we have compared the $\alpha_{ox}$ distribution 
of the X-ray selected NLSy1 from Grupe et al. (2010) (left-hand shaded histogram in Figure~\ref{figure=aox}) with that
of the {\it INTEGRAL} NLSy1 (right-hand shaded histogram). A Kolmogorov-Smirnov (K-S) test results in a probability $P = 0.057$ of a random result,
indicating that the distributions are not significantly different. We find two extreme cases
of a very X-ray-to-optical loud source IRAS 15091-2107 ($\alpha_{ox}$ = 0.9) and an X-ray quiet source Swift~J2127.4+5654. However,
the strong Galactic extinction for both galaxies may not guarantee a correct estimate of the dereddened flux; 
indeed, as shown by Grupe et al. (2010), the reddening correction influences the final measurements and distributions.

A small black hole mass of 10$^{7}$ M$_{\odot}$ is residing in 1H 0323+342 which is a gamma-ray source,
detected by {\it Fermi}, with a spectral energy distribution typical of BL Lac objects. This is at odds with 
the evidence that radio-loud AGN are generally associated with the most massive black holes and 
are known to reside mostly in elliptical galaxies (e.g. McLure \& Jarvis 2004, Dunlop et al. 2003). 
Interestingly, at least three sources in our sample are ellipticals (IRAS~15091-2107, IGR~J16385-2057 and IGR~J19378-0617)  
characterized by black hole masses $<$ 10$^{7}$ M$_{\odot}$ and, they are radio sources.
In Table~\ref{sam} we report the radio fluxes at 1.4 or 5 GHz.
The strongest radio sources in our sample are 1H 0323+342 and NGC 5506.
However, if we calculate the X-ray radio-loudness parameter R$_{X}$ $\equiv$ L$_{\nu}$(R)/L(2-10 keV),
only 1H 0323+342 can strictly be considered as radio-loud (log R$_{X}$=-2.7) if we assume the more conservative
radio-quiet versus radio-loud boundary as defined in Panessa et al. (2007). Otherwise, if we assume the limit set by
Terashima \& Wilson (2003) to log R$_{X}$ $=$ -4.5, 
source in the sample should be considered as radio-loud, except for IGR~J16385-2057, IGR~J19378-0617 and Swift~J2127.4+5654. 
The radio-loudness parameters
obtained from the low angular resolution integrated flux should, however, be taken with caution, since
it has been shown that observations at high angular resolution can resolve $>$ 40\% of the larger scale emission,
resulting in a much reduced core radio luminosity and a lower radio-loudness parameter,
as shown from interferometric measurements of NGC 4051 which provide 
log R$_{X}$ $=$ -5.8, suggesting that the central AGN is very radio-quiet (Giroletti \& Panessa 2009). 
Considering the 11 NLSy1 with radio measurements, the fraction of radio-loud sources in our sample is likely $\sim$ 9\%,
consistently with Komossa (2007) who found that 7\% of the NLSy1 galaxies in their sample were radio-loud and
only 2.5\% ``very'' radio loud (R $>$ 100). This result also confirms the search for NLSy1 in the FIRST survey
(the Faint Images of the Radio Sky at 20 cm survey; Becker et al. 1995), where a decline of 
the NLSy1 fraction is observed toward high R-values, i.e.,
the NLSy1 fraction falls down to 10\% in moderately strong radio AGNs (Zhou et al. 2006).

\section{Conclusions}

Hard X-ray NLSy1 look very heterogeneous in their broad-band properties, displaying a broad range of hard X-ray (20-100 keV)
photon indeces, flatly distributed from quite flat ($\sim$ 1.3) to very steep ($\sim$ 3.6) slopes. At these energies no clear separation
between the BLSy1 and NLSy1 slopes distributions is found, although the hard X-ray selection may introduce a bias against
faint steep spectrum sources. Steeper photon indeces for NLSy1 are instead found when considering the broad-band spectra.
In only one source, Swift~J2127.4+5654, a high energy cut-off is measured at a relatively low value (E$_{cut-off}$ $\sim$ 50 keV),
and the reflection fraction constrained to R=1.0$^{+0.5}_{-0.4}$, confirming previous measurements (Malizia et al. 2008, Miniutti et al. 2009). 
Apparently, the hard X-ray selection is not efficient in detecting strong soft X-ray NLSy1 as, indeed, 
only one source, IGR~J19378-0617, shows a dominant and strongly variable soft X-ray
component. The presence of fully or partially covering absorption is quite frequently measured in the spectra 
(in four out of ten sources analyzed in this work). 
When the spectral quality is good (i.e., with XMM-\textit{Newton} data), we almost always detect the presence of an Fe line 
(showing from narrow to moderately broad profiles). 
A proper determination of the reflection strength is limited by the non simultaneity of the X-ray and hard X-ray observations. 
This is particularly true in this class of sources where X-ray variability (in flux and spectrum) is strong, 
both at short and long timescales, as here demonstrated by the XMM-\textit{Newton} light curves and in those sources where multiple 
observations were available. {\it INTEGRAL} and BAT hard X-rays average fluxes
are consistent, with a maximum flux variation of $\sim$ 30\%. 
Overall, the average X-ray spectrum of NLSy1 looks more similar to 
the BLSy1 spectrum when the sources are selected at hard X-ray rather than soft X-rays (e.g., {\it ROSAT}). 

We estimate that the fraction of NLSy1 in the hard X-ray sky is about 10-15\% of the 
type 1 AGN population, in agreement with the optically selected sample fraction.
The black hole mass distribution of NLSy1 selected in hard X-rays peaks at 10$^{7}$ M$_{\odot}$ 
and the Eddington ratios distribution peaks at 10$^{-2}$, suggesting that hard X-ray NLSy1 
occupy the lower tail of Eddington ratios distribution of NLSy1. 
When data are available, NLSy1 in our sample are radio emitting objects, though not radio-loud,
except for 1H 0323+342 which is a blazar-like NLSy1.

\section*{Acknowledgments}
We thank the referee, Dirk Grupe, for his useful comments and suggestions which helped us improving our paper.
We thank Alessandro Maselli for his precious contribution to the {\it Swift}/UVOT data analysis.
FP and RL acknowledge support by {\it INTEGRAL} ASI I/033/10/0 and
ASI/INAF I/009/10/0. GM thanks the Spanish Ministry of Science and Innovation for support
through grants AYA2009-08059 and AYA2010-21490-C02-02.

{99}


\begin{thebibliography}{99}

\bibitem[Abdo et al.(2009)]{2009ApJ...707L.142A} Abdo, A.~A., et al.\ 2009, \apjl, 707, L142 
\bibitem[Ajello et al.(2008)]{2008ApJ...673...96A} Ajello, M., et al.\ 2008, \apj, 673, 96 
\bibitem[Arnaud(1996)]{1996ASPC..101...17A} Arnaud, K.~A.\ 1996, Astronomical Data Analysis Software and Systems V, 101, 17 
\bibitem[Barthelmy et al.(2005)]{2005SSRv..120..143B} Barthelmy, S.~D., et al.\ 2005, \ssr, 120, 143 
\bibitem[Bassani et al.(2006)]{2006ApJ...636L..65B} Bassani, L., et al.\ 2006, \apjl, 636, L65 
\bibitem[Becker et al.(1995)]{1995ApJ...450..559B} Becker, R.~H., White, R.~L., \& Helfand, D.~J.\ 1995, \apj, 450, 559 
\bibitem[Becker et al.(1991)]{1991ApJS...75....1B} Becker, R.~H., White, R.~L., \& Edwards, A.~L.\ 1991, \apjs, 75, 1 
\bibitem[Bennett et al.(2003)]{2003ApJS..148....1B} Bennett, C.~L., et al.\ 2003, \apjs, 148, 1 
\bibitem[Bianchi et al.(2009)]{2009A&A...495..421B} Bianchi, S., Guainazzi, M., Matt, G., Fonseca Bonilla, N., \& Ponti, G.\ 2009a, \aap, 495, 421 
\bibitem[Bianchi et al.(2009)]{2009A&A...501..915B} Bianchi, S., Bonilla, N.~F., Guainazzi, M., Matt, G., \& Ponti, G.\ 2009b, \aap, 501, 915 
\bibitem[Bird et al.(2010)]{2010ApJS..186....1B} Bird, A.~J., et al.\ 2010, \apjs, 186, 1 
\bibitem[Bird et al.(2007)]{2007ApJS..170..175B} Bird, A.~J., et al.\ 2007, \apjs, 170, 175 
\bibitem[Boller et al.(2003)]{2003MNRAS.343L..89B} Boller, T., Tanaka, Y., Fabian, A., Brandt, W.~N., Gallo, L., Anabuki, N., Haba, Y., \& Vaughan, S.\ 2003, \mnras, 343, L89 
\bibitem[Boller et al.(2002)]{2002MNRAS.329L...1B} Boller, T., et al.\ 2002, \mnras, 329, L1 
\bibitem[Boller et al.(1996)]{1996A&A...305...53B} Boller, T., Brandt, W.~N., \& Fink, H.\ 1996, \aap, 305, 53 
\bibitem[Brandt et al.(1997)]{1997MNRAS.285L..25B} Brandt, W.~N., Mathur, S., \& Elvis, M.\ 1997, \mnras, 285, L25 
\bibitem[Burrows et al.(2005)]{2005SSRv..120..165B} Burrows, D.~N., et al.\ 2005, \ssr, 120, 165 
\bibitem[Butler et al.(2009)]{2009ApJ...698..502B} Butler, S.~C., et al.\ 2009, \apj, 698, 502 
\bibitem[Comastri(2000)]{2000NewAR..44..403C} Comastri, A.\ 2000, \nar, 44, 403 
\bibitem[Condon et al.(1998)]{1998AJ....115.1693C} Condon, J.~J., Cotton, W.~D., Greisen, E.~W., Yin, Q.~F., Perley, R.~A., Taylor, G.~B., \& Broderick, J.~J.\ 1998, \aj, 115, 1693 
\bibitem[Cusumano et al.(2010)]{2010A&A...510A..48C} Cusumano, G., et al.\ 2010, \aap, 510, A48 
\bibitem[Deluit(2004)]{2004A&A...415...39D} Deluit, S.~J.\ 2004, \aap, 415, 39 
\bibitem[Denney et al.(2009)]{2009ApJ...702.1353D} Denney, K.~D., et al.\ 2009, \apj, 702, 1353 
\bibitem[de Rosa et al.(2008)]{2008A&A...483..749D} de Rosa, A., Bassani, L., Ubertini, P., Panessa, F., Malizia, A., Dean, A.~J., \& Walter, R.\ 2008, \aap, 483, 749 
\bibitem[Dewangan \& Griffiths(2005)]{2005ApJ...625L..31D} Dewangan, G.~C., \& Griffiths, R.~E.\ 2005, \apjl, 625, L31 
\bibitem[Dietrich et al.(2005)]{2005ApJ...623..700D} Dietrich, M., Crenshaw, D.~M., \& Kraemer, S.~B.\ 2005, \apj, 623, 700 
\bibitem[Dunlop et al.(2003)]{2003MNRAS.340.1095D} Dunlop, J.~S., McLure, R.~J., Kukula, M.~J., Baum, S.~A., O'Dea, C.~P., \& Hughes, D.~H.\ 2003, \mnras, 340,1095 
\bibitem[Fabian et al.(2004)]{2004MNRAS.353.1071F} Fabian, A.~C., Miniutti, G., Gallo, L., Boller, T., Tanaka, Y., Vaughan, S., \& Ross, R.~R.\ 2004, \mnras, 353, 1071 
\bibitem[Foschini et al.(2009)]{2009AdSpR..43..889F} Foschini, L., Maraschi, L., Tavecchio, F., Ghisellini, G., Gliozzi, M., \& Sambruna, R.~M.\ 2009, Advances in Space Research, 43, 889 
\bibitem[Gallimore et al.(2006)]{2006AJ....132..546G} Gallimore, J.~F., Axon, D.~J., O'Dea, C.~P., Baum, S.~A., \& Pedlar, A.\ 2006, \aj, 132, 546 
\bibitem[Gallo(2006)]{2006MNRAS.368..479G} Gallo, L.~C.\ 2006, \mnras, 368, 479 
\bibitem[Gehrels et al.(2004)]{2004ApJ...611.1005G} Gehrels, N., et al.\ 2004, \apj, 611, 1005 
\bibitem[George et al.(2000)]{2000ApJ...531...52G} George, I.~M., Turner, T.~J., Yaqoob, T., Netzer, H., Laor, A., Mushotzky, R.~F., Nandra, K., \& Takahashi, T.\ 2000, \apj, 531, 52 
\bibitem[Gierli{\'n}ski \& Done(2004)]{2004MNRAS.349L...7G} Gierli{\'n}ski, M., \& Done, C.\ 2004, \mnras, 349, L7 
\bibitem[Giroletti \& Panessa(2009)]{2009ApJ...706L.260G} Giroletti, M., \& Panessa, F.\ 2009, \apjl, 706, L260 
\bibitem[Goldwurm et al.(2003)]{2003A&A...411L.223G} Goldwurm, A., et al.\ 2003, \aap, 411, L223 
\bibitem[Goodrich(1989)]{1989ApJ...342..224G} Goodrich, R.~W.\ 1989, \apj, 342, 224 
\bibitem[Grupe et al.(2010)]{2010ApJS..187...64G} Grupe, D., Komossa, S., Leighly, K.~M., \& Page, K.~L.\ 2010, \apjs, 187, 64 
\bibitem[Grupe \& Mathur(2004)]{2004ApJ...606L..41G} Grupe, D., \& Mathur, S.\ 2004, \apjl, 606, L41 
\bibitem[Grupe(2004)]{2004AJ....127.1799G} Grupe, D.\ 2004, \aj, 127, 1799 
\bibitem[Guainazzi et al.(2010)]{2010MNRAS.406.2013G} Guainazzi, M., Bianchi, S., Matt, G., Dadina, M., Kaastra, J., Malzac, J., \& Risaliti, G.\ 2010, \mnras, 406, 2013 
\bibitem[]{}Halpern, J. P. 2006, ATel, 847
\bibitem[Hasinger et al.(2000)]{2000NewAR..44..497H} Hasinger, G., Lehmann, I., Schmidt, M., Gunn, J.~E., Schneider, D.~P., Giacconi, R., Tr{\"u}mper,  J., \& Zamorani, G.\ 2000, \nar, 44, 497 
\bibitem[Hill et al.(2004)]{2004SPIE.5165..217H} Hill, J.~E., et al.\ 2004, \procspie, 5165, 217 
\bibitem[Kaspi et al.(2000)]{2000ApJ...533..631K} Kaspi, S., Smith, P.~S., Netzer, H., Maoz, D., Jannuzi, B.~T., \& Giveon, U.\ 2000, \apj, 533, 631 
\bibitem[Komossa(2007)]{2007ASPC..373..719K} Komossa, S.\ 2007, The Central Engine of Active Galactic Nuclei, 373, 719 
\bibitem[Komossa\& Xu(2007)]{2007ApJ...667L..33K} Komossa, S., \& Xu, D.\ 2007, \apjl, 667, L33 
\bibitem[Landi et al.(2010)]{2010MNRAS.403..945L} Landi, R., Bassani, L., Malizia, A., Stephen, J.~B., Bazzano, A., Fiocchi, M., \& Bird, A.~J.\ 2010, \mnras, 403, 945 
\bibitem[Landi et al.(2005)]{2005A&A...441...69L} Landi, R., Malizia, A., \& Bassani, L.\ 2005, \aap, 441, 69
\bibitem[Leighly(1999)]{1999ApJS..125..317L} Leighly, K.~M.\ 1999, \apjs, 125, 317 
\bibitem[Longinotti et al.(2003)]{2003A&A...410..471L} Longinotti, A.~L., Cappi, M., Nandra, K., Dadina, M., \& Pellegrini, S.\ 2003, \aap, 410, 471 
\bibitem[Malizia et al.(2010)]{2010MNRAS.408..975M} Malizia, A., Bassani, L., Sguera, V., Stephen, J.~B., Bazzano, A., Fiocchi, M., \& Bird, A.~J.\ 2010, \mnras, 408, 975 
\bibitem[Malizia et al.(2009)]{2009MNRAS.399..944M} Malizia, A., Stephen, J.~B., Bassani, L., Bird, A.~J., Panessa, F., \& Ubertini, P.\ 2009, \mnras, 399, 944 
\bibitem[Malizia et al.(2008)]{2008MNRAS.389.1360M} Malizia, A., et al.\ 2008, \mnras, 389, 1360 
\bibitem[Manchanda(2006)]{2006AdSpR..38.1387M} Manchanda, R.~K.\ 2006, Advances in Space Research, 38, 1387 
\bibitem[Marconi et al.(2008)]{2008ApJ...678..693M} Marconi, A., Axon, D.~J., Maiolino, R., Nagao, T., Pastorini, G., Pietrini, P., Robinson, A., \& Torricelli, G.\ 2008, \apj, 678, 693 
\bibitem[Masetti et al.(2010)]{2010A&A...519A..96M} Masetti, N., et al.\ 2010, \aap, 519, A96 
\bibitem[Masetti et al.(2009)]{2009A&A...495..121M} Masetti, N., et al.\ 2009, \aap, 495, 121 
\bibitem[Masetti et al.(2008)]{2008A&A...482..113M} Masetti, N., et al.\ 2008, \aap, 482, 113 
\bibitem[Masetti et al.(2007)]{2007A&A...470..331M} Masetti, N., et al.\ 2007, \aap, 470, 331 
\bibitem[Masetti et al.(2006)]{2006A&A...459...21M} Masetti, N., Morelli, L., Palazzi, E., et al.\ 2006, \aap, 459, 21
\bibitem[Mason et al.(2001)]{2001A&A...365L..36M} Mason, K.~O., et al.\ 2001, \aap, 365, L36 
\bibitem[Mathur et al.(2001)]{2001NewA....6..321M} Mathur, S., Kuraszkiewicz, J., \& Czerny, B.\ 2001, \na, 6, 321 
\bibitem[McLure \& Jarvis(2004)]{2004MNRAS.353L..45M} McLure, R.~J., \& Jarvis, M.~J.\ 2004, \mnras, 353, L45 
\bibitem[Miller et al.(2010)]{2010MNRAS.403..196M} Miller, L., Turner, T.~J., Reeves, J.~N., Lobban, A., Kraemer, S.~B., \& Crenshaw, D.~M.\ 2010, \mnras, 403, 196 
\bibitem[Miniutti et al.(2009)]{2009MNRAS.398..255M} Miniutti, G., Panessa, F., de Rosa, A., Fabian, A.~C., Malizia, A., Molina, M., Miller, J.~M., \& Vaughan, S.\ 2009, \mnras, 398, 255 
\bibitem[Molina et al.(2009)]{2009MNRAS.399.1293M} Molina, M., et al.\ 2009, \mnras, 399, 1293 
\bibitem[Moretti et al.(2004)]{2004SPIE.5165..232M} Moretti, A., et al.\ 2004, \procspie, 5165, 232 
\bibitem[Nagar et al.(2002)]{2002A&A...391L..21N} Nagar, N.~M., Oliva, E., Marconi, A., \& Maiolino, R.\ 2002, \aap, 391, L21 
\bibitem[Nandra et al.(1997)]{1997ApJ...477..602N} Nandra, K., George, I.~M., Mushotzky, R.~F., Turner, T.~J., \& Yaqoob, T.\ 1997, \apj, 477, 602 
\bibitem[Niko{\l}ajuk et al.(2009)]{2009MNRAS.394.2141N} Niko{\l}ajuk, M., Czerny, B., \& Gurynowicz, P.\ 2009, \mnras, 394, 2141 
\bibitem[Osterbrock \& De Robertis(1985)]{1985PASP...97.1129O} Osterbrock, D.~E., \& De Robertis, M.~M.\ 1985, \pasp, 97, 1129 
\bibitem[Osterbrock \& Pogge(1985)]{1985ApJ...297..166O} Osterbrock, D.~E., \& Pogge, R.~W.\ 1985, \apj, 297, 166 
\bibitem[Panessa et al.(2008)]{2008A&A...483..151P} Panessa, F., et al.\ 2008, \aap, 483, 151 
\bibitem[Panessa et al.(2007)]{2007A&A...467..519P} Panessa, F., Barcons, X., Bassani, L., Cappi, M., Carrera, F.~J., Ho, L.~C., \& Pellegrini, S.\ 2007, \aap, 467, 519 
\bibitem[Ponti et al.(2010)]{2010MNRAS.406.2591P} Ponti, G., et al.\ 2010, \mnras, 406, 2591 
\bibitem[Ponti et al.(2006)]{2006MNRAS.368..903P} Ponti, G., Miniutti, G., Cappi, M., Maraschi, L., Fabian, A.~C., \& Iwasawa, K.\ 2006, \mnras, 368, 903 
\bibitem[Reynolds(1997)]{1997MNRAS.286..513R} Reynolds, C.~S.\ 1997, \mnras, 286, 513 
\bibitem[Risaliti et al.(2009)]{2009ApJ...700L...6R} Risaliti, G., Young, M., \& Elvis, M.\ 2009, \apjl, 700, L6 
\bibitem[]{}Rodriguez-Ardila, A., Pastoriza, M. G., Donzelli, C. J. 2000, ApJS, 126, 63
\bibitem[Roming et al.(2005)]{2005SSRv..120...95R} Roming, P.~W.~A., et al.\ 2005, \ssr, 120, 95 
\bibitem[Shemmer et al.(2008)]{2008ApJ...682...81S} Shemmer, O., Brandt, W.~N., Netzer, H., Maiolino, R., \& Kaspi, S.\ 2008, \apj, 682, 81 
\bibitem[Sulentic et al.(2000)]{2000ApJ...536L...5S} Sulentic, J.~W., Zwitter, T., Marziani, P., \& Dultzin-Hacyan, D.\ 2000, \apjl, 536, L5 
\bibitem[Tanaka et al.(2004)]{2004PASJ...56L...9T} Tanaka, Y., Boller, T., Gallo, L., Keil, R., \& Ueda, Y.\ 2004, \pasj, 56, L9 
\bibitem[Telezhinsky et al.(2010)]{2010A&A...522A..68T} Telezhinsky, I., Eckert, D., Savchenko, V., Neronov, A., Produit, N., \& Courvoisier, T.~J.-L.\ 2010, \aap, 522, A68 
\bibitem[Terashima et al.(2009)]{2009PASJ...61S.299T} Terashima, Y., et al.\ 2009, \pasj, 61, 299 
\bibitem[]{} Terashima, Y.~\& Wilson, A.~S.\ 2003, \apj, 583, 145 
\bibitem[Tueller et al.(2010)]{2010ApJS..186..378T} Tueller, J., et al.\ 2010, \apjs, 186, 378 
\bibitem[Tueller et al.(2009)]{2009AAS...21330103T} Tueller, J., Markwardt, C.~B., Skinner, G.~K., Baumgardner, W.~H., \& Swift BAT Survey Team 2009, Bulletin of the American Astronomical Society, 41, 269 
\bibitem[Turner et al.(2007)]{2007A&A...475..121T} Turner, T.~J., Miller, L., Reeves, J.~N., \& Kraemer, S.~B.\ 2007, \aap, 475, 121 
\bibitem[Turner et al.(1999)]{1999ApJ...526...52T} Turner, T.~J., George, I.~M., \& Netzer, H.\ 1999, \apj, 526, 52 
\bibitem[Ubertini et al.(2003)]{2003A&A...411L.131U} Ubertini, P., et al.\ 2003, \aap, 411, L131 
\bibitem[Ulvestad et al.(1995)]{1995AJ....109...81U} Ulvestad, J.~S., Antonucci, R.~R.~J., \& Goodrich, R.~W.\ 1995, \aj, 109, 81 
\bibitem[Uttley et al.(2004)]{2004MNRAS.347.1345U} Uttley, P., Taylor, R.~D., McHardy, I.~M., Page, M.~J., Mason, K.~O., Lamer, G., \& Fruscione, A.\ 2004, \mnras, 347, 1345 
\bibitem[V{\'e}ron-Cetty \& V{\'e}ron(2006)]{2006A&A...455..773V} V{\'e}ron-Cetty, M.-P., \& V{\'e}ron, P.\ 2006, \aap, 455, 773 
\bibitem[Wandel et al.(1999)]{1999ApJ...526..579W} Wandel, A., Peterson, B.~M., \& Malkan, M.~A.\ 1999, \apj, 526, 579 
\bibitem[Wang \& Lu(2001)]{2001A&A...377...52W} Wang, T., \& Lu, Y.\ 2001, \aap, 377, 52 
\bibitem[Whalen et al.(2006)]{2006AJ....131.1948W} Whalen, D.~J., Laurent-Muehleisen, S.~A., Moran, E.~C., \& Becker, R.~H.\ 2006, \aj, 131, 1948 
\bibitem[Williams et al.(2002)]{2002AJ....124.3042W} Williams, R.~J., Pogge, R.~W., \& Mathur, S.\ 2002, \aj, 124, 3042 
\bibitem[Winkler et al.(2003)]{2003A&A...411L...1W} Winkler, C., et al.\ 2003, \aap, 411, L1 
\bibitem[Winter et al.(2008)]{2008ApJ...674..686W} Winter, L.~M., Mushotzky, R.~F., Tueller, J., \& Markwardt, C.\ 2008, \apj, 674, 686 
\bibitem[Yuan et al.(2008)]{2008ApJ...685..801Y} Yuan, W., Zhou, H.~Y., Komossa, S., Dong, X.~B., Wang, T.~G., Lu, H.~L., \& Bai, J.~M.\ 2008, \apj, 685, 801 
\bibitem[Zhou et al.(2007)]{2007ApJ...658L..13Z} Zhou, H., et al.\ 2007, \apjl, 658, L13 
\bibitem[Zhou et al.(2006)]{2006ApJS..166..128Z} Zhou, H., Wang, T., Yuan, W., Lu, H., Dong, X., Wang, J., \& Lu, Y.\ 2006, \apjs, 166, 128 
\bibitem[Zhou \& Wang(2002)]{2002ChJAA...2..501Z} Zhou, H.-Y., \& Wang, T.-G.\ 2002, \cjaa, 2, 501 


\end{thebibliography}
\end{document}